\documentclass[runningheads]{llncs}
\usepackage[title]{appendix}
\usepackage{amsmath,amssymb,amsfonts}
\usepackage{algorithmic}
\usepackage{graphicx}
\usepackage{textcomp}
\usepackage{xcolor}
\usepackage{colortbl}
\usepackage{multirow}

\usepackage{wrapfig}

\usepackage{mathtools}
\usepackage{adjustbox}
\usepackage{subfigure}

\usepackage{algorithmic}
\usepackage{algorithm}
\usepackage[misc]{ifsym}

\newcommand{\ie}{{\em i.e., }}
\newcommand{\eg}{{\em e.g., }}

\def\BibTeX{{\rm B\kern-.05em{\sc i\kern-.025em b}\kern-.08em
    T\kern-.1667em\lower.7ex\hbox{E}\kern-.125emX}}

\begin{document}

\title{Network Anatomy and Real-Time Measurement of Nvidia GeForce NOW Cloud Gaming}
\titlerunning{Cloud Gaming Network Anatomy and Real-Time Measurement}

\author{Minzhao Lyu\inst{1} \and Sharat Chandra Madanapalli\inst{2} \and \\ Arun Vishwanath\inst{2} \and Vijay Sivaraman\inst{1}}
\authorrunning{M. Lyu \textit{et al.}}
\institute{University of New South Wales, Sydney, Australia \\ \email{ \{minzhao.lyu,vijay\}@unsw.edu.au}  \and Canopus Networks Pty Ltd \\ \email{ \{sharat,arun\}@canopusnet.com} }

\maketitle 

\begin{abstract}
	Cloud gaming, wherein game graphics is rendered in the cloud and streamed back to the user as real-time video, expands the gaming market to billions of users who do not have gaming consoles or high-power graphics PCs. Companies like Nvidia, Amazon, Sony and Microsoft are investing in building cloud gaming platforms to tap this large unserved market. However, cloud gaming requires the user to have high-bandwidth and stable network connectivity -- whereas a typical console game needs about 100-200 kbps, a cloud game demands minimum 10-20 Mbps. This makes the Internet Service Provider (ISP) a key player in ensuring the end-user's good gaming experience.

	In this paper we develop a method to detect Nvidia’s GeForce NOW cloud gaming sessions over their network infrastructure, and measure associated user experience. In particular, we envision ISPs taking advantage of our method to
	provision network capacity at the right time and in the right place to support growth in cloud gaming at the right experience level; as well as identify the role of contextual factors such as user setup (browser vs app) and connectivity type (wired vs wireless) in performance degradation. We first present a detailed anatomy of flow establishment and volumetric profiles of cloud gaming sessions over multiple platforms, followed by a method to detect gameplay and measure key experience aspects such as latency, frame rate and resolution via real-time analysis of network traffic. The insights and methods are also validated in the lab for XBox Cloud Gaming platform. We then implement and deploy our method in a campus network to capture gameplay behaviors and experience measures across various user setups and connectivity types which we believe are valuable for network operators.
	\end{abstract}

\section{Introduction}

Cloud gaming, also known as Game-as-a-Service (GaaS), allows users to play graphics-intensive games without having to own expensive hardware such as gaming consoles (\eg PlayStation and X-Box) and PCs equipped with high-end graphics cards. Instead, game graphics is rendered on powerful cloud-hosted servers, and the resulting video is streamed instantaneously to the end-user's device. Though cloud gaming is still nascent, it is estimated to grow 58.75\% each year to become a \$22.53 billion industry by 2030 \cite{Economist2022,Economist2023,GitnuxBlog}. Entities such as Nvidia, Amazon, Sony, and Microsoft, as well as a slew of smaller companies, have commercial cloud gaming offerings in the market already.

A significant barrier to the growth of cloud gaming is the high bandwidth that it demands from the network. Whereas a typical game (\eg shooting game like Call-of-Duty or sports game like FIFA) played on a console or PC requires only a few hundred kilobits-per-second (kbps) from the network \cite{SCMadanapalliPAM2022}, a cloud game demands two orders of magnitude more, namely tens of Megabits-per-second (Mbps). If such higher bandwidth (coupled with low latency) is not consistently available, gaming experience becomes frustrating due to drop in frames-per-second (fps), poor resolution, and jerky movements. Internet Service Providers (ISPs) often bear the brunt of this frustration, leading to complaints, churn, and reputational damage.

ISPs have historically overprovisioned their networks with respect to peak aggregate volumes, agnostic to the application mix. This is ceasing to be cost-effective as access speeds increase, because elastic applications (like large downloads) can grab unconstrained amounts of network bandwidth at any instant, degrading performance for experience-sensitive applications (like cloud gaming). ISPs are therefore looking for ways to distinguish (and potentially segregate) the latter -- as one example, Comcast recently announced \cite{comcast} it will support low latency forwarding in its cable broadband network for application streams marked by specific content providers (such as Apple and Nvidia). ISPs are therefore seeking to gain visibility and control into which content is worthy of preferential bandwidth or latency treatment, custom tune their network provisioning by locale in an on-demand manner to retain cost-efficiency, and continuously monitor user experience on cloud gaming and other sensitive applications to protect brand reputation.

To the best of our knowledge, little research has been done for ISPs to measure the prevalence of, and experience on, cloud gaming over their network infrastructure with fine visibility into user setups and functionalities of traffic flows, so that they can configure their networks (\eg via network APIs and slices) for guaranteed user experience. Prior works have analysed website browsing \cite{NWehnerPER2021}, video streaming \cite{BSpangBS2020,SLiuMWUT2023,sharma2023estimating}, live video \cite{SCMadanapalliIWQoS2021}, console/PC gaming \cite{SCMadanapalliPAM2022}, \textit{etc.}, but not cloud gaming. Existing studies of cloud gaming have considered video processing delays on client devices/cloud servers \cite{HIqbalMACS2021}, energy consumption on mobile devices \cite{SBhuyanMACS2022}, differentiating cloud game RTP flows \cite{JKyISCC2023} from UDP traffic, impact of wireless/edge network conditions on gameplay RTP flows \cite{JKyCNSM2022,JKyTNSM2023,MCarrascosaCC2022}, which are not of relevance to ISPs who seek insights to better manage their networks.
	
Our objective in this work is to develop practical methods that can be deployed by ISPs to gain fine-grained visibility into cloud gaming behaviors and experience, so they can be actively involved in, rather than be cut out of, managing service delivery quality. We have chosen to primarily focus on Nvidia's cloud gaming platform, GeForce NOW (GFN), in this paper for a few reasons: (a) it is currently recognized as the leader in the global cloud gaming market \cite{ClougGamingMarket}; (b) it supports a rich collection of multiplayer games from all major game publishers like Steam, Ubisoft, and Epic Games; (c) it has local servers hosted in the geography where our experiments were conducted; (d) the research team has access to the personnel operating the GFN servers for the region, making it easier to seek clarifications and corroborate findings; and (e) it is accessible broadly, both via the browser and a bespoke app, on desktops (\ie macOS and Windows) and mobile devices (\ie android and iOS), thereby providing a rich set of conditions under which it can be studied. 
The obtained insights and developed methods are also preliminarily validated in our lab for XBox Cloud Gaming that together with GeForce NOW dominate over 90\% of market share globally \cite{JRietveld2023} and in Australia.

Our \textit{first contribution} (\S\ref{sec:TrafficAnalysis}) reveals the detailed anatomy of GeForce NOW cloud gaming. We identify the establishment of various flows, and benchmark their volumetric patterns, associated with gameplay management, user input, and audio/video streaming, while highlighting the differences between the user playing on a browser versus the app. We also identify attributes in the traffic that aid in user experience estimation, such as sequence numbers for measuring latency and marker packets or stochastic patterns in packet payload sizes for detecting frame boundaries.

Our \textit{second contribution} (\S\ref{sec:inNetworkDetection}) develops a practical network traffic analysis method to detect cloud gaming sessions and measure gameplay experience. We use a stateful mapping mechanism that tracks service domains accessed by active user flows so as to detect the start of a cloud gaming session, as well as identify the user setup (\ie operating system and software agent type). We then categorize the gameplay flows as gaming management, user input, and video streaming based on volumetric attributes. Network operators can therefore prioritize certain gameplay flows for guaranteed user experience such as mapping user inputs into low-latency network slices (\eg URLLC) while assigning gaming video to high-bandwidth slices. Finally, we derive user experience measures, including client-platform latency (between the gamer device and cloud platform), video frame rate, and video graphic resolution.

For our \textit{third contribution} (\S\ref{sec:evaluationAndFieldInsights}), we implement our methods as a prototype and deploy it in a University campus network\footnote{We have obtained ethics clearance (UNSW Human Research Ethics Advisory Panel approval number HC211007) which allows us to analyze campus network traffic to infer application usage behaviors. Note that user identities remain anonymous -- no attempt is made to extract or reveal any personal user information, and all results presented are aggregated across the campus.} with on-premise student housing. We evaluate the accuracy of our method via gameplay in the lab on-campus. We then collect data in the wild, and present some interesting insights obtained over the course of a month spanning 362 hours of playtime. We found about 36\% of playtime was via browsers, which were all in low resolution (Standard Definition) - probably because the browser is not as optimized as the app. The laptop app accounted for 51\% of playtime, and was for the most part in High Definition, though interestingly it tended to reduce frame rate from 60 fps to 30 fps more often, most likely so it could preserve the higher video resolution. The mobile app accounted for 14\% of playtime, and was largely in high resolution. Knowing the mix of user setups will help ISPs tune and troubleshoot their networks to uplift user experience in a cost-efficient manner.

\begin{table*}[t!]
	\caption{Popular cloud gaming platforms on the market and their specifications.}
	\label{tab:specification}
	\small
	\begin{tabular}{|l|l|l|l|l|l|}
		\hline
		\rowcolor[rgb]{ .906,  .902,  .902}  \textbf{Platform} & \textbf{Operator} & \textbf{Operating System} & \textbf{Soft. Agent}  \\ \hline
		GeForce NOW \cite{GeForceNOW} & Nvidia  & Windows, macOS, iOS, Android &  app, browser  \\ \hline		
		XBox Cloud Gaming \cite{XBoxCloudGame} & Microsoft  &  Windows, iOS, android, XBox    &  app, browser  \\ \hline
		PlayStation PLUS  \cite{PlayStationNOW} &   Sony &    Windows, PS4, PS5  &  app   \\ \hline
		Luna \cite{Luna} & Amazon   & Windows, macOS, android, iOS   &  app, browser \\ \hline
	\end{tabular}
\end{table*}

\section{Cloud Gaming Background \& Workflow}\label{sec:Development}
In this section, we provide an overview of current development in cloud gaming and highlight the challenges that arise for Internet service providers (\S\ref{sec:development}). We then discuss typical operational process of gameplay on a cloud gaming platform (\S\ref{sec:OperationalProcess}).

\subsection{Development of Cloud Gaming}\label{sec:development}
In order to provide gamers with the opportunity to play graphic-intensive games on their everyday PCs without the need to invest in high-end devices, leading companies in graphic processing (\eg Nvidia), cloud services (\eg Amazon), and gaming operations (\eg Sony) have embarked on the development of the ``cloud gaming'' business model, also referred to as Game-as-A-Service. This model moves on-device gaming processing to the powerful cloud compute clusters. 
Under this model, gamers can subscribe to the service and gain access to a dedicated cloud platform that supports a wide range of graphics-demanding games. The gameplay scenes are then rendered in real-time on the cloud servers and streamed to the user's device.

The popular cloud gaming platforms operated by the above-mentioned tech giants are listed in Table~\ref{tab:specification}, with their specifications including operator, supported operating systems (OS), and software agents. It is clear that all of the four platforms are accessible on major OSes and support both console applications and browsers, except that XBox Cloud Gaming does not support macOS and only available via its console application.

However, such operational mode shifts the requirement of running high-end games from expensive graphic processing and computing hardware on user device to the network's capability of streaming high-resolution live video, along with other essential gaming requirements such as low latency and jitter. This can result in significant data consumption, often exceeding several tens of Megabits per second.
In comparison to traditional online games that primarily exchange lightweight flags consisting of user inputs and server responses, which typically require only a few kilobits per second, cloud gaming places unprecedented burdens on carrier networks. If not properly managed, the increased network demands can lead to customer frustration and even prompt users to switch network providers. 
Recognizing the emerging challenges for ISPs, as an early step, we undertake a study that focuses on the network traffic characteristics of a representative cloud gaming platform, GeForce NOW with the aim to detect cloud gameplay and measure gamer experience.

\subsection{Operational Process of Cloud Game Sessions}\label{sec:OperationalProcess}
Cloud gaming platforms serve as intermediaries that connect user devices and individual game servers.
These platforms receive input from gamers including mouse movement, keyboard input, and upstream audio, and perform graphic and gaming computations on their behalf. Real-time gameplay scenes are streamed back to users via video and audio services.
Therefore, a typical cloud gaming process involves two types of sessions that are in charge of platform administration and actual gameplay, respectively.

\begin{figure}[t!]
	{\includegraphics[width=\textwidth]{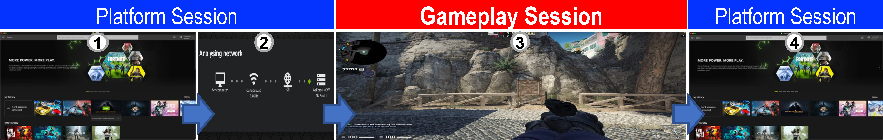}}
	\caption{A typical cloud gaming process.}
	\label{fig:ExampleProcess}
\end{figure}

We now walk through an example play of a popular first-person shooting game (\ie Counter-Strike: Global Offensive or its abbreviation CS:GO) on GeForce NOW platform.
As shown in the leftmost window of Fig.~\ref{fig:ExampleProcess}, after logging into the cloud gaming platform via either console application or browser, the gamer is directed to the ``\textbf{platform session}'', during which the gamer can browse available games and choose the one to play. In addition, some of the graphical settings such as resolution and frame rate are also options to be set by the gamer during the platform session.
Once a gameplay is about to start, as shown by the second left-most window in Fig.~\ref{fig:ExampleProcess}, the optimal cloud server for this gameplay is selected via a set of network measurements for key performance metrics, such as latency and throughput between the gamer device and candidate clusters.

After initial setup, the ``\textbf{gameplay session}'' begins, such as the CS:GO shooting gameplay visually depicted in Fig.~\ref{fig:ExampleProcess}. 
As will be soon discussed in \S\ref{sec:TrafficAnalysis}, the gameplay session places significant demand on network throughput and latency, which directly impact the quality of experience (QoE) for cloud gaming. 
After exiting a gameplay session, a gamer returns to the platform session (the rightmost window of Fig.~\ref{fig:ExampleProcess}) to select/start next gameplay or finish the cloud game session.

\section{Cloud Gaming Network Traffic Characterization}\label{sec:TrafficAnalysis}
We now delve into the network traffic characteristics of gameplay on the GeForce NOW platform obtained from our labeled traffic traces captured in our lab environment (\S\ref{sec:Dataset}). We begin by discussing the anatomy of service flows (\S\ref{sec:UsageOfServiceFlow}) in both platform and gameplay sessions, and then focusing on the profile of critical flows in gameplay sessions (\S\ref{sec:GamePlaySessionFlows}). We also characterize XBox Cloud Gaming to demonstrate the general applicability of our insights/methods for platforms sharing similar underlying technologies (Appendix~\S\ref{sec:AppendixXBoxFlowProfile}).

\subsection{Dataset}\label{sec:Dataset}
The GeForce NOW cloud gaming platform is available via both console applications and browsers on PCs (\ie macOS and Windows) and mobile devices (\ie iOS and android). Therefore, we capture packet trace files (PCAP) during cloud gaming sessions of a real-time shooting game (\ie CS:GO) and a massive multiplayer role-play game (\ie Path of Exile) on the above user setups, which can be categorized as either \textbf{desktop console application}, \textbf{mobile console application}, or \textbf{browser}. 
In what follows, we primarily focus on the insights from gameplays on the console application installed in a macOS desktop, while the differences in other supported setups (\ie Chrome browser in macOS, console application in Windows, Chrome browser in Windows, console application in android, and Safari browser in iOS) are also discussed throughout the section. To validate our obtained insights, we further collected traffic traces of 20 cloud game sessions for each type of user setups. The 20 cloud gaming sessions for each setup type cover three commonly available graphic resolutions from full high-definition (FHD), high definition (HD), to standard definition (SD), with either 30fps or 60fps video frame rates.

\subsection{Anatomy of Service Flows}\label{sec:UsageOfServiceFlow}
\begin{figure*}[t!]
	{\includegraphics[width=\textwidth]{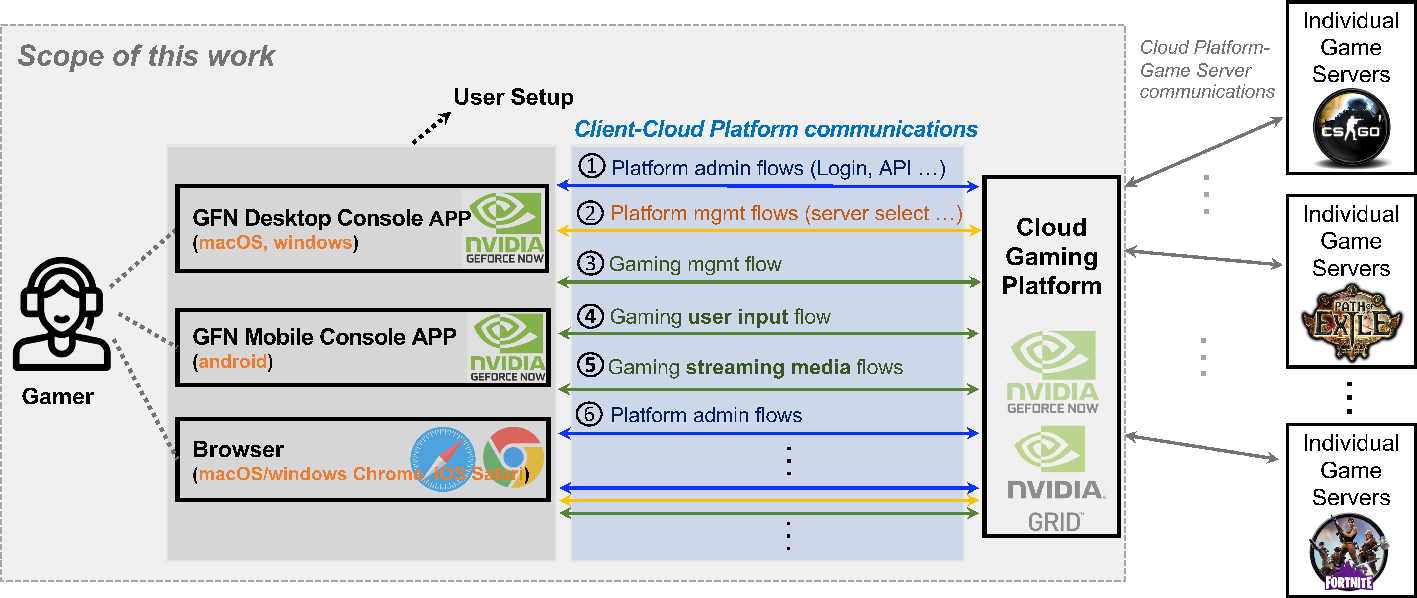}}
	\caption{Anatomy of cloud gaming communications via GeForce NOW (GFN) platform.}
	\label{fig:CommunicationAnatomy}
\end{figure*}

As visually shown in Fig.~\ref{fig:CommunicationAnatomy}, we first look at the anatomy of network communications between a gamer device and the cloud gaming platform as obtained from our analytical results. There are three types of flows that collectively serve a cloud gaming session, namely for \textbf{platform administration}, \textbf{platform management}, and \textbf{gameplay} that are illustrated as blue, yellow, and green arrows in Fig.~\ref{fig:CommunicationAnatomy}, respectively.

Specifically, once a user opens the cloud gaming console application or browser, \ie entering a platform session, a series of administration flows (\ie \textcircled{1} in Fig.~\ref{fig:CommunicationAnatomy}) are initialized for administrative support, such as content management system (CMS), API utilities, login portals, and account management. After the user selects a game to play, a number of platform management flows (\ie \textcircled{2} in Fig.~\ref{fig:CommunicationAnatomy}) are started for network diagnosis and cloud server selection. Those flows are mapped to the second screenshot in Fig.~\ref{fig:ExampleProcess} and hold critical roles in the successful initiation of the subsequent gameplay. 
During actual gameplay sessions, three types of flows are initialised, each serving a specific purpose. Firstly, there are gaming management flows (\ie \textcircled{3}) responsible for controlling the gameplay session (\eg measuring run-time latency) and exchanging metadata (\eg coordinating interactions between the client and the platform). 
In addition, there are specific gameplay flows dedicated to user input (from mouse, keyboard and microphone) and streaming media contents (\ie downstream video and audio), annotated as \textcircled{4} and \textcircled{5} respectively
After a gameplay session, the platform administration flows are restarted as the user is returned to the platform session.

We now give a representative example of the above-discussed service flow usage as a time-series plot in Fig.~\ref{fig:FlowProfileGFN}.
In this example, we played two games on the macOS PC console application. 
Our user activity could be separated into six stages. We logged into the cloud game platform and selected our first game (\ie CS:GO) to play in stage 1; played the CS:GO game in stage 2; returned to the platform in stage 3; entered the login and character selection page of our second game (\ie Path of Exile) in stage 4; entered the world of Path of Exile in stage 5, and finished our cloud gaming play in stage 6. 
We note that each of the six activity stages are either platform or gameplay session as defined in Fig.~\ref{fig:ExampleProcess}, thus, in Fig.~\ref{fig:FlowProfileGFN}, they are annotated by blue or red labels for the two session types, respectively.

In Fig.~\ref{fig:FlowProfileGFN}, we could see the timespans of all relevant flows that are with identical color indicating their flow types as consistent in Fig~\ref{fig:CommunicationAnatomy}. 
The median throughput of each flow is indicated by the line thickness in Fig.~\ref{fig:FlowProfileGFN}.
A representative collection of flows are labeled (as y-axis ticks) in the format of simplified service prefix (extracted from SNI or DNS records) and identifiable port numbers. The labels for other flows are not included for readability. 
Now we discuss the details per flow type.

\subsubsection{Platform Administration Flows}\label{sec:platformAdmin}
\begin{figure*}[t!]
	\centering
	{\includegraphics[width=0.95\textwidth]{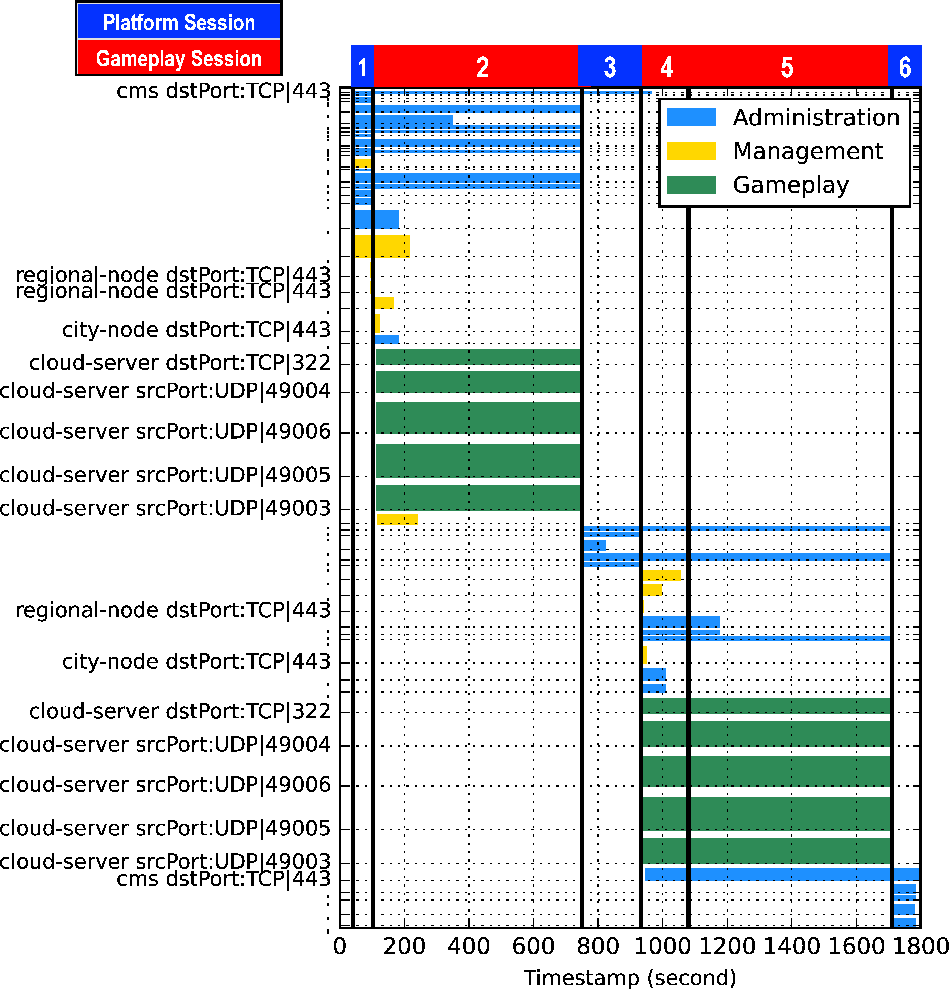}}
	\caption{Flow profiles of example gameplay via GeForce NOW desktop console application. Service prefixes and port numbers of representative flows are shown by their respective y-ticks.  Throughputs of flows are indicated by their bar thicknesses (normalised by logarithmic functions).}
	\label{fig:FlowProfileGFN}
\end{figure*}

Platform administration flows are shown as blue lines in Fig.~\ref{fig:FlowProfileGFN}. They are all sent to the service port $TCP|443$. After decoding packet headers, we confirm that they are all HTTPS flows toward the provider root domains (\ie \textit{nvidia.com}, \textit{geforcenow.com}, and \textit{geforce.com}) for administrative services as indicated in their subdomain prefixes, such as content management system (CMS) \cite{CMS} (\textit{cms}) and frontend APIs (\textit{gx-target-experiments-frontend-api}). Depending on their service purposes, some of the flows (\eg \textit{login} and \textit{userstore}) are only active during the platform session, whereas others (\eg \textit{cms} and \textit{events}) remain active during the subsequent gameplay sessions.
As for their volumetric profiles, all of them are having small (or even negligible) amount of volume usage, \ie less than several Megabytes.

Upon comparing the platform administration flows across different user setups, we have observed that the sessions via Chrome browser (an example is shown in Appendix Fig.~\ref{fig:FlowProfileChrome}) has most of the service flows seen in our discussed example, except for \textit{cms} and \textit{als} which are related to high-performant graphics. Besides, unlike PC setups, cloud game sessions via android mobile console application have limited usage of platform administrative flows (an example is given in Appendix Fig.~\ref{fig:FlowProfileAndroid}) that seem to only cover essential services such as \textit{login}, \textit{event}, and \textit{userstore}.

\subsubsection{Platform Management Flows}\label{sec:PlatformManagement}
As depicted by the yellow horizontal bars in Fig~\ref{fig:FlowProfileGFN}, the platform management flows exhibit a relatively lower quantity compared to the platform administration flows. They are all HTTPS flows sent to service port $TCP|443$ of the GeForce NOW cloud server clusters, which are all associated with the cloud cluster domain \textit{nvidiagrid.net}. 

Unlike the above-discussed administration flows that provide support to the user administration on the platform, such as login, events, and user store, the platform management flows service critical tasks relating to the delivery of subsequent gameplay session. 
In stage 1 of Fig.~\ref{fig:FlowProfileGFN}, we observe the first management (yellow) flow directed towards the subdomain prefix \textit{gfnpc.api.entitlement-prod}, which grants the client access to the GeForce NOW production system. This is followed by multiple flows toward subdomain prefixes in the format of \textit{server\_[location]\_pnt}, which serve the purpose of cloud server selection by measuring network performance metrics between the user and various available vantage points.
As indicated by the service prefixes in Fig.~\ref{fig:FlowProfileGFN} from top to down, the selection process progresses from the regional node to the city node and ultimately to each individual cloud server.

Given the indispensable roles of platform management flows, \ie system access and server selection, there is no significant difference that can be observed across different user setups, except that four flows toward the service prefix \textit{img} of the production system domain \textit{nvidiagrid.net} are seen in platform sessions on android mobile but not on PC setups.

\subsubsection{Gameplay Session Flows}\label{sec:GameSession}
Once a suitable cloud server is successfully selected by the platform management flows, the actual gameplay session is started.
As depicted by the green lines in stage 2 of Fig.~\ref{fig:FlowProfileGFN}, five gameplay session flows are initiated for the CS:GO gameplay. This observation remains consistent for all gameplays via console applications on both mobile and PC devices. Similarly, in stage 4 and 5 of Fig.~\ref{fig:FlowProfileGFN}, the same combination of five gameplay session flows can be observed for the Path of Exile gameplay.

Specifically, the first gameplay session flow is always directed to destination service port $TCP|322$ on the cloud server. This is followed by four Real-Time Transport Protocol (RTP) flows originating from client ports $UDP|49003$, $UDP|49004$, $UDP|49005$, and $UDP|49006$ toward dynamically selected service ports on the cloud server. According to Nvidia Support \cite{NividiaPortMapping}, the four client ports are assigned for downstream audio, upstream audio, downstream video, and user input, respectively. 
We believe that the dynamic selection of service ports is performed by the first TCP flow for gaming management purpose, as illustrated in Fig.~\ref{fig:CommunicationAnatomy}.

Notably, the flows originating from $UDP|49005$, responsible for downstream video, consistently consumed larger amounts of bandwidth than all other gameplay session flows. In the 10-minute CS:GO gameplay (stage 1), this resulted in downstream video data transfers of 1422MB, while in the Path of Exile gameplay (stages 4 and 5), it reached 1961MB. As for the flows for downstream audio, upstream audio, and upstream user input, they consumed several MB, several tens of MB, and several tens of MB data transfers, respectively.

Similar insights are observed for gameplay session flows across console applications on both mobile and desktop devices. However, in browser-based gameplay, their gaming management flows is delivered using WebRTC protocol toward the service port $TCP|49100$ \cite{NividiaPortWebRTC}, as shown by Fig.~\ref{fig:FlowProfileChrome} in the Appendix. Additionally, gameplays on browsers utilize a single UDP flow to carry both downstream media contents and upstream user input, in contrast to the separate flows seen in console applications.

Given that the user gaming experience on cloud platforms is predominantly influenced by the gameplay session flows, in what follows, we will focus on examining their flow-level volumetric profiles and packet statistics.

\begin{figure*}[t!]
	{\includegraphics[width=\textwidth]{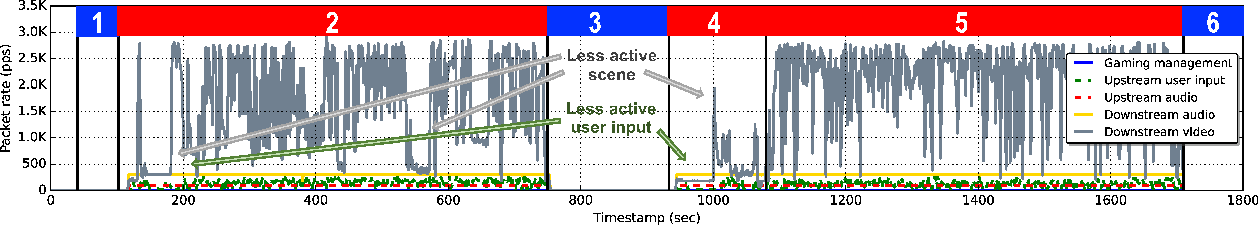}}
	\caption{Volumetric profiles of gameplay session flows (the green lines in Fig.~\ref{fig:FlowProfileGFN}).}
	\label{fig:FlowProfileGameplay}
\end{figure*}

\subsection{Profile of Gameplay Session Flows}\label{sec:GamePlaySessionFlows}

We now focus on gameplay session flows (represented by the green lines in Fig.~\ref{fig:FlowProfileGFN}) that are critical to the cloud gaming experience. 
For flows that are active during a gameplay session, we analyze their volumetric profiles (\S\ref{sec:FlowVolumetric}); benchmark their bandwidth consumptions across different levels of graphic quality including frame rate and resolution (\S\ref{sec:BandwidthConsumption}); and explore how video frame rate could be identified by packet size patterns of flows carrying downstream video data (\S\ref{sec:FrameRate}).

\subsubsection{Flow Volumetric Profile}\label{sec:FlowVolumetric}
For our example gameplay discussed in Fig.~\ref{fig:FlowProfileGFN}, the inbound and outbound packet rates of the five gameplay session flows, including gameplay management, upstream user input, upstream audio, downstream audio, and downstream video, are presented in a time-series plot shown in Fig.~\ref{fig:FlowProfileGameplay}.

First, the \textbf{gameplay management flow}, represented by the inbound and outbound packets towards the service port $TCP|322$, exhibits a low packet rate of one pair of packets every two seconds (\ie 0.5pps), which may appear negligible in the plot. It can serve as a reliable indicator for measuring network latencies between the user and the cloud platform by tracking the sequence and acknowledgment numbers in TCP packet headers.

Second, the upstream \textbf{user input flow} for keyboard and mouse actions exhibits a packet rate ranging from 11 to 267 pps, depending on the user's activity level. As depicted by the green lines in Fig.~\ref{fig:FlowProfileGameplay}, the baseline packet rate (11pps) occurs when the user is actively moving the mouse or typing on the keyboard, such as during the matching phase of a shooting game (beginning of stage 2 in Fig.~\ref{fig:FlowProfileGameplay}) or the opening cinematic of a role-play game (beginning of stage 4). It is worth noting that the outbound and inbound packet rates of this flow are often equal. 
Regarding the throughput of the user input flow, the outbound direction exhibits minimum and maximum values of 1k and 82kbps, respectively, which are approximately ten times larger than those observed in the inbound direction.

Third, the two flows for \textbf{upstream and downstream audio} both have constant packet rates and throughputs in their respective directions. 
Specifically, the upstream audio flow exhibits a packet rate of 100pps with a throughput of 15kbps, while the downstream audio flow has a packet rate of 300pps with a throughput of 37kbps. 
These values remain consistent regardless of the status of the input/output voice, such as being muted, at low volume, or at high volume, which we thoroughly tested during our experiments.
In their opposite directions (\eg inbound direction for upstream audio flow), both of them constantly have 2 pps packet rate with 156bps throughput.

\begin{wraptable}{L}{0.6\textwidth}
	\vspace{-8mm}
	\caption{Representative \textit{Peak} bandwidth consumption of downstream video flows across graphic resolutions and frame rates during active gameplay. The \textbf{\color{green}green}, \textbf{\color{yellow}yellow}, and \textbf{\color{red}red} cells are for full high-definition ({\color{green}FHD}), high-definition ({\color{yellow}HD}), and standard-definition ({\color{red}SD}), respectively.}
	\label{tab:bitrate}
	\small
	\begin{tabular}{|l|l|l|l|l|l|l|l|l|l|l|l|}
		\hline
		&\cellcolor{green}\textbf{FHD} & \cellcolor{yellow} \textbf{HD} &\cellcolor{pink}\textbf{SD} \\ \hline
		
		\textbf{60FPS} & \cellcolor{green} 23  -- 35Mbps   & \cellcolor{yellow} 15 -- 21Mbps  & \cellcolor{pink}$<=$13Mbps   \\ \hline
		
		\textbf{30FPS} & \cellcolor{green}  15 -- 22Mbps   &  \cellcolor{yellow} 9 -- 13Mbps    & \cellcolor{pink} $<=$8Mbps    \\ \hline
	\end{tabular}
\end{wraptable}

The most bandwidth consuming flow type, \ie \textbf{downstream video flows}, exhibit a packet rate that ranges from approximately 300pps during less active scenes (as shown in Fig.~\ref{fig:FlowProfileGameplay}) to 3000pps. 
The corresponding bandwidth consumptions for these packet rates are 3Mbps and 34Mbps, respectively.
It is important to note that during active gameplay periods, such as stage 5 in Fig.~\ref{fig:FlowProfileGameplay}), the bandwidth consumption mostly remains at the upper bound level.
On the other hand, the low packet rate is observed during inactive periods, which often coincide with the inactivity of the user input flows. However, there are exceptions during static scenes with frequent user inputs, such as the login scene of a gameplay.

Similar patterns in terms of packet rates and bandwidth consumption are observed for gameplay session flows across both desktop and mobile console applications. However, for gameplay sessions on browsers, there is only one gameplay session flow that combines all four flows for user input and media. In this case, the volumetric pattern of the flow is primarily dominated by the downstream video content.

We note that the numerical results presented earlier were specific to the video configuration with 60fps for the frame rate and a resolution of 1920x1080 (FHD). However, users have the flexibility to choose from a wide range of frame rates and resolutions, either statically or dynamically adjusted based on the network conditions during gameplay. In the following analysis, we will discuss the bandwidth consumption of downstream video flows for different graphic configurations.

\subsubsection{Bandwidths of Downstream Video Flows across Video Configurations}\label{sec:BandwidthConsumption}

Frame rate and graphic resolution are configurable parameters for a cloud gameplay. To analyze the bandwidth consumption for downstream video flows, we manually selected various available frame rates (30 or 60fps) and graphic resolutions (ranging from 1920x1200 to 1024x768). For browser sessions where only one gameplay flow is present, we also measured the bandwidth consumption as it is primarily influenced by downstream video content.
As discussed for Fig.~\ref{fig:FlowProfileGameplay}, the packet rate and throughput for downstream video flows stay at a peak range during active gameplay scenarios. In Table~\ref{tab:bitrate}, we report the observed peak bandwidth consumption of those downstream video flows under each different graphic configuration.

Each cell in Table~\ref{tab:bitrate} is color-coded into three groups based on the resolution types including full high-definition (FHD), high-definition (HD), and standard definition (SD). Green cells represent the ideal FHD graphic quality, while yellow and red cells indicate less optimal graphic configurations (HD and SD) that may result in a subpar gaming experience. 
In general, lower frame rates and coarse-grained graphic resolutions always result in lower bandwidth consumption, both in active and less active scenarios. 
It is worth noting that by examining the peak bandwidth consumption of video flows and the current frame rate (which could be inferred from packet size patterns and will be discussed soon), it is possible for an ISP to infer the current graphic resolution of an active cloud gameplay by its user.

 \begin{figure*}[t!]
	\begin{center}
		\mbox{
			\subfigure[MacOS PC console app - 60FPS.]{
				{\includegraphics[width=0.52\textwidth]{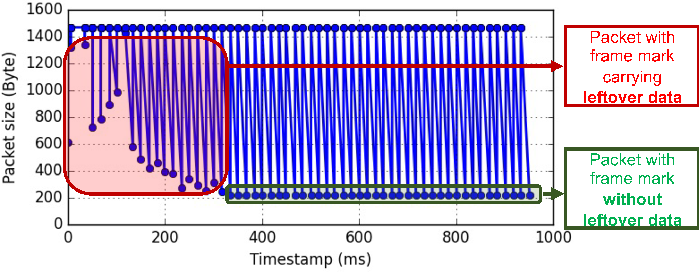}}\quad
				\label{fig:VideoFlowPktProfileMacGFN}
			}
			\subfigure[Windows PC console app - 30FPS.]{
				{\includegraphics[width=0.38\textwidth]{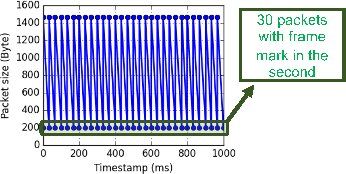}}\quad
				\label{fig:VideoFlowPktProfileWindowsGFN}
			}
		}
		\mbox{
			\subfigure[Multiple packets.]{
				{\includegraphics[width=0.32\textwidth]{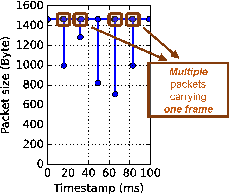}}\quad
				\label{fig:VideoFlowPktProfileWindowsGFNMultiplePackets}
			}		
			\hspace{6mm}	
			\subfigure[Windows PC Chrome browser - 30FPS.]{
				{\includegraphics[width=0.48\textwidth]{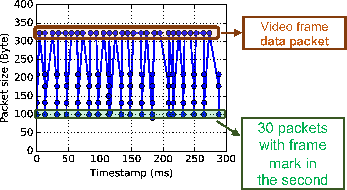}}\quad
				\label{fig:VideoFlowPktProfileChrome}
			}
		}
		\caption{Packet payload sizes in representative downstream video flows (each packet is represented as a blue dot) as time-series plots.}
		\label{fig:videoFrame}
	\end{center}
\end{figure*}

\subsubsection{Packet Size Patterns in Downstream Video Flows across Frame Rates}\label{sec:FrameRate}

Identifying the number of packets with a frame marker set in the RTP header of downstream video flows is a straightforward method to benchmark the current frame rate. In RTP flows, the frame marker indicates the completion of the currently transmitted video frame \cite{RTPRFC}. By counting the packets with the frame marker in the downstream video flow, we can accurately determine the frame rate of a cloud gameplay. 
However, in a high-speed network, this method requires decoding RTP packets that could introduce non-negligible overheads (\eg via multiple sequential packet parsers each decoding one packet layer). Besides, frame marks may not always be set correctly or being encrypted, makes RTP headers not decodable.
For a lightweight and robust method, we have observed certain patterns in the packet sizes of downstream video flows to determine the frame rate without decoding the Ethernet/IP/UDP/RTP headers. This approach ensures resilience even when frame marks are not available. 
Now we look at several example downstream video flows shown in Fig.~\ref{fig:videoFrame}.

In general, a video frame is carried by two types of RTP packets. The first type carries all or the majority of video frame data and has a fixed large packet size (\ie 1466 bytes in console applications). The second type consists of smaller-sized packets (\eg 216 bytes) that carry frame markers, indicating the completion of a frame transmission.
During gameplay sessions, the number of data packets required to transmit a video frame is dynamically adjusted based on the amount of video data in the frame. This can range from one packet to multiple packets, as visually depicted in Fig.~\ref{fig:VideoFlowPktProfileWindowsGFNMultiplePackets}. Importantly, there is always one small-sized marker packet indicating the end of each frame, which is highlighted by the green box in Fig.~\ref{fig:VideoFlowPktProfileMacGFN}.
These frame marker packet may have larger sizes (still smaller than the size of data packets) to accommodate any remaining video data from the previous packets. This is indicated by the red box in Fig.~\ref{fig:VideoFlowPktProfileMacGFN}.

We observed a consistent pattern in the downstream video flows, where the number of packet groups (comprising several data packets followed by one marker packet) aligns perfectly with the current frame rate. An example of this pattern can be seen in Fig.~\ref{fig:VideoFlowPktProfileWindowsGFN}, where we observe 30 packet groups within one second for a frame rate of 30fps. By analyzing the size patterns of these packet groups, we can accurately identify the frame rate being received by a cloud gamer in real-time.

In contrast to gameplay via console applications, the scenario for browsers is slightly different as there is only one RTP flow responsible for carrying downstream video, audio, and user input data. As illustrated in Fig.~\ref{fig:VideoFlowPktProfileChrome}, each vertically aligned group of packets consists of several data packets corresponding to a video frame, followed by a frame marker packet, and three additional packets for downstream audio, upstream audio, and user input. Despite this difference, the consistent pattern observed in each packet group via browsers still aligns with the delivery of video frames.

\section{Gameplay Detection and Experience Measurement}\label{sec:inNetworkDetection}

\begin{figure}[t]
	\centering
	\includegraphics[width=0.8\textwidth]{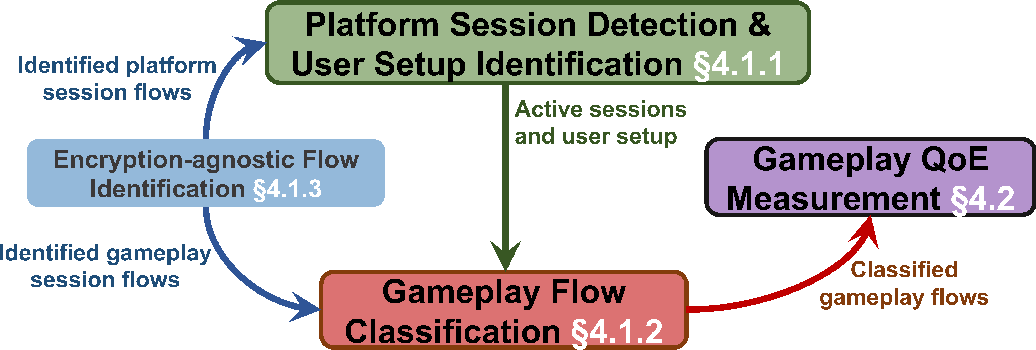}
	\caption{The abstract view of our cloud gameplay detection and experience measurement framework in \S\ref{sec:inNetworkDetection}.}
	\label{fig:detectionMethod}
\end{figure}

In this section, building upon the insights gained from \S\ref{sec:TrafficAnalysis}, we present the development of our network traffic analysis framework that detects cloud gaming sessions, identifies user setups, and continuously measures the quality-of-experience (QoE) metrics of each cloud gameplay session, as illustrated in Fig.~\ref{fig:detectionMethod}.  We also discuss the generalizability of our method to other cloud gaming platforms and briefly discuss limitations of our QoE metrics (\S\ref{sec:generalizabilityAndLimitation}).

\subsection{Detecting Cloud Game Sessions}
As previously shown in Fig.~\ref{fig:ExampleProcess} and \ref{fig:FlowProfileGameplay}, a cloud gameplay typically goes through two types of sessions, namely platform session (for platform administration, game browsing, and cloud server selection) and gameplay session.
From our analysis, platform sessions exhibit identical usage of service flows in terms of user setups, which are important for a network operator to better understand their customer segments and potential causes of QoE degradation. The flows in gameplay sessions directly determine users' cloud game experience.

Note that our method uses the identification of flows directed towards specific service domains based on the Server Name Indication (SNI) field in the SSL headers, which may not be available due to the potential adoption of encrypted-SNI (ESNI). Therefore, we have also devised an encryption-agnostic approach (Appendix \S\ref{sec:flowIdentification}) to overcome this challenge in the near future.

We now present our method to detect both platform and gameplay sessions.

\subsubsection{Platform Sessions and User Setup Identification}\label{sec:detectingPlatformSession}

User enters platform sessions either in the initial phase after launching to the cloud game platform (\eg stage 1 in Fig.~\ref{fig:FlowProfileGFN}) or in a subsequent phase after finishing each gameplay session (\eg stage 3 and 6 in Fig.~\ref{fig:FlowProfileGFN}).
As discussed in \S\ref{sec:UsageOfServiceFlow}, when a platform session begins, a series of HTTPS flows are initiated towards different service prefixes of the cloud gaming domains. These HTTPS flows contain service names in the server name indication (SNI) \cite{SNI} fields of their SSL handshakes, which precede the encrypted application data communication.
Therefore, we detect platform sessions by monitoring flows toward each service domain, extracted from their respective SNI records. 

After analyzing our traffic traces across all user setups, we categorize the flows into two types including core services and setup-specific services.
Core service flows are always initiated during the starting phase of platform sessions, regardless of the user setup type. There are also service flows specific to certain user setups. For instance, flows directed toward \textit{login.nvidia} are always active, while flows toward \textit{play.nvidia} only occur in platform sessions via browsers.
Additionally, we have observed that the majority of core service flows follow a sequential order, starting from user login and progressing to server selection. In contrast, the occurrence of setup-specific service flows often exhibits randomness in their sequence.

\begin{figure}[t!]
	{\includegraphics[width=\textwidth]{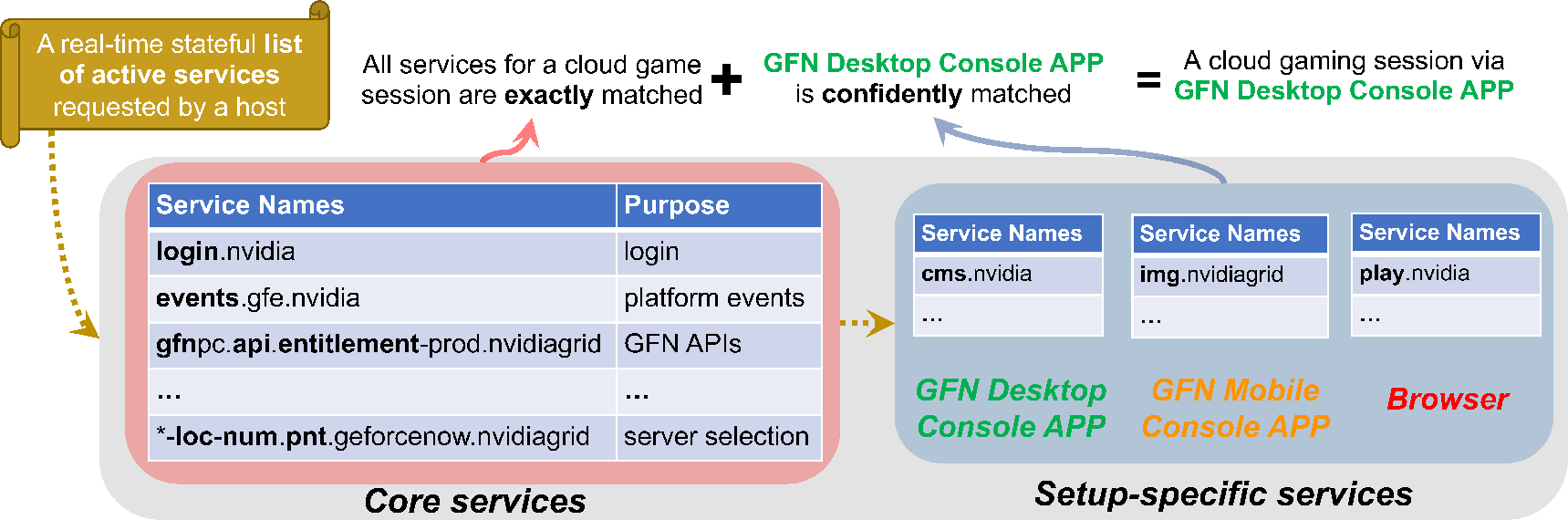}}
	\caption{Process for the detection of platform sessions and identification of user setups.}
	\label{fig:platformSessionDetection}
\end{figure}

As shown in Fig.~\ref{fig:platformSessionDetection}, we have developed a codebook correlation of domain names to detect the start of a cloud gaming play (through core services) and identify the user setup type (\ie desktop console application, mobile console application, or browser). A wildcard match of domains in the corresponding table will trigger a successful detection. 
For example, as demonstrated by the top portion of Fig.~\ref{fig:platformSessionDetection}, the exact match of flows in core service table triggers the successful detection of a cloud gaming session. Additionally, a confident match in the desktop console application table (compared to the other tables) determines that the cloud game session is being played on a desktop console. Please note that for simplicity, we have provided only a snippet of our matching tables rather than the entire list.

\subsubsection{Gameplay Sessions and Gameplay Flow Classification}\label{sec:detectingGameplaySession}
As a recap from \S\ref{sec:GamePlaySessionFlows}, after starting a gameplay on the cloud platform, one SSL-encrypted TCP gaming management flow and several (one for browser and four for console application) UDP flows are started for media and user input.
Cloud gaming sessions from console applications have their TCP gaming management flows directed to service port numbers $TCP|322$, while browsers use port $TCP|49100$. 
Irrespective of the user setup types, all gaming management flows have their service name (extracted from SNI) following a consistent pattern of \textit{a-b-c-d}.\textit{pnt.nvidiagrid.net}. It is important to note that the \textit{a-b-c-d} in this pattern represents the IP address \textit{a.b.c.d} of a cloud server assigned to a gameplay session, which is also the server IP for the subsequent gameplay UDP flows.

\begin{figure*}[t!]
	\centering
	{\includegraphics[width=0.9\textwidth]{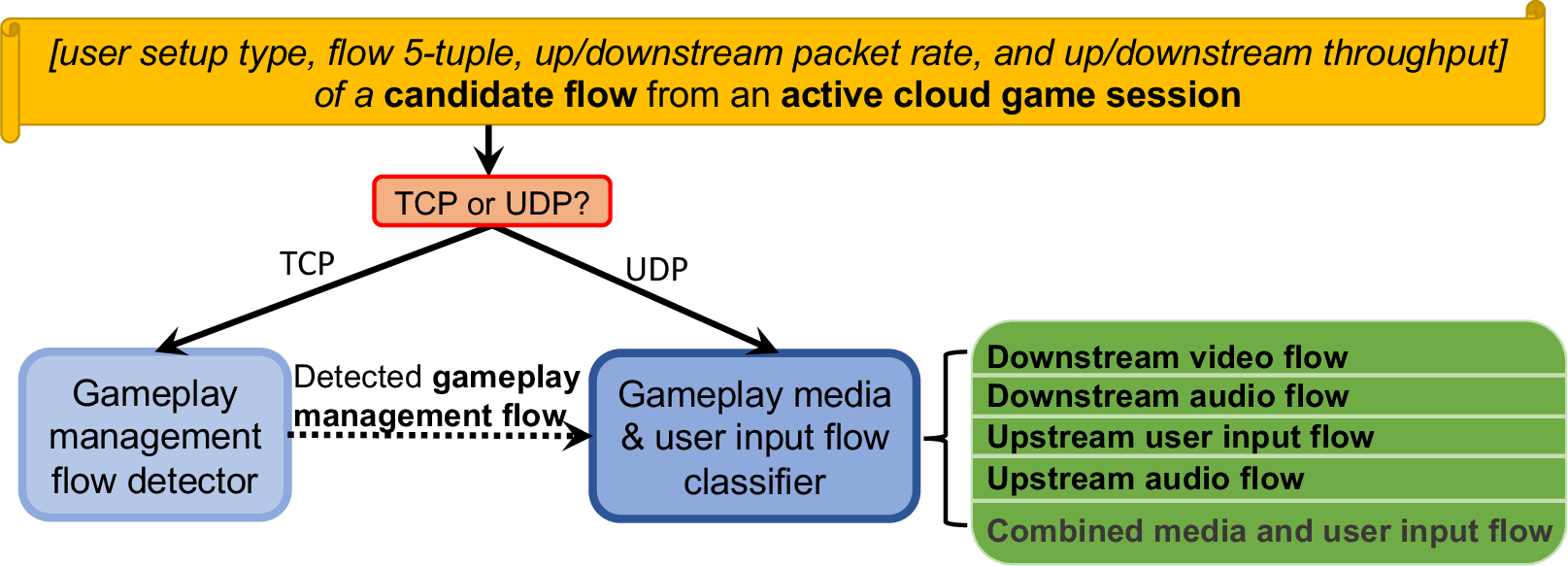}}
	\caption{Process for the classification of gameplay session flows, wherein the models can be either obtained using generalized machine-learning algorithms or heuristic criteria for a specific cloud gaming platform.}
	\label{fig:gameplaySessionDetection}
\end{figure*}

We devise our method to identify gameplay session flows based on their service names and five-tuples for active users detected in the platform sessions (\S\ref{sec:detectingPlatformSession}). Subsequently, the gameplay session flows are then classified for their purposes as defined in \S\ref{sec:GamePlaySessionFlows}, based on their user setup types, flow five-tuple, and volumetric profiles (\ie packet rate and throughput). 
The generalized classification process is shown in Fig.~\ref{fig:gameplaySessionDetection}. The models in this process can be obtained either using machine learning algorithms on standardized input attributes (shown as the yellow banner in Fig.~\ref{fig:gameplaySessionDetection}) that can be directly applied to cloud gaming platforms implemented using similar mechanisms, or heuristically derived for a certain platform (\eg GeForce NOW) according to the specific profiles of its gameplay flows.
In this work, based on our insights obtained for GFN flow characteristics, we developed an automatic training script to derive flow classification criteria using ground-truth traffic traces (\ie PCAP files).
We also provide flow charts showing the process for GFN and XBox Cloud Gaming with heuristically simplified classification criteria in Appendix~\ref{sec:AppendixGameplayFlowClassification} Fig.~\ref{fig:gameplaySessionDetection-GFN} and \ref{fig:gameplaySessionDetection-XBox}, respectively.

The first step of the classification process, depicted in the left blue model in Fig.~\ref{fig:gameplaySessionDetection}), is to \textbf{classify gameplay management flows}. For GeForce NOW, a candidate TCP flow (identified by its service name) sent to port $TCP|322$ or $TCP|49100$ from an active user are for console application or browser, respectively. As indicated by the dashed arrow, the five-tuple (including client and server IP addresses) of the gameplay management flow will be stored in a runtime database for the detection of subsequent gameplay UDP flows that are initiated shortly afterward (\eg within 0.5s).

The second step, illustrated in the right blue model of Fig.~\ref{fig:gameplaySessionDetection}, is to \textbf{classify gameplay media and user input flows}. 
In the case of browser-based GFN sessions, a gameplay UDP flow can either carry media and user input data or provide support for the STUN WebRTC service, which can be differentiated based on its packet rate. Specifically, a STUN WebRTC service flow has its packet rate less than 2pps while a combined media and user input flow has over hundreds of packets per second.
For gameplay UDP flows in console application-based GFN sessions, they can serve different purposes, including downstream video, downstream audio, upstream audio, or user input. Classification of these flows is based on their volumetric profiles, specifically the packet rate and throughput, in both inbound and outbound directions. From our training process on GFN traffic traces, criteria on volumetric attributes are obtained including $>5$Mbps inbound throughput for downstream video flows, matched inbound and outbound packet rates (\ie $\Delta <=10$pps) for user input flow, and inbound or outbound dominant packet rates (\ie $\Delta >10$pps) for downstream or upstream voice flows, respectively.
It may be argued that the functionality of a gameplay UDP flow from GFN console application can be easily determined by its client port number as discussed in \S\ref{sec:GameSession}. However, the presence of network address translation (NAT) in ISP networks often obfuscates the client port numbers of users, rendering them unreliable in comparison to our method that is based on volumetric profiles.

In \S\ref{sec:QoEMetrics}, we will discuss the continuous monitoring of the downstream video flows via console applications, the flows carrying media with user input via browsers, and gaming management flows as an effective approach to infer user experience per cloud gameplay session.

\subsubsection{Encryption-Agnostic Service Flow Identification}\label{sec:flowIdentification}
The recent proposals of SNI encryption techniques have sparked discussions in the industry and could potentially be adopted in the near future. If that happens, current network analysis relying on SNI signatures will become ineffective.
To tackle this anticipated issue, we develop an encryption-agnostic technique (the blue module in Fig.~\ref{fig:detectionMethod}) as an enhancement to our existing flow detection using service names extracted from the SNI field.
This enhancement allows us to identify flows with their service names associated with platform and gameplay sessions by analyzing packet payload sizes, without the need to inspect SSL headers.

\begin{table}[t!]
	\centering
	\caption{Packet payload sizes (in byte) after three-way handshakes of gameplay TCP management flows for different user setups (OS and software agent) observed in both \textbf{\color{purple}upstream $\uparrow$} and \textbf{\color{blue}downstream $\downarrow$} directions.}
	\small
	\begin{tabular}{|l|l|}
		\hline
		\rowcolor[rgb]{ .906,  .902,  .902} 	\textbf{OS}     & \textbf{dstPort numbers and payload sizes across software agent types} \\ \hline
		Windows  &    \textit{$TCP|322$ (APP)} or \textit{$TCP|49100$ (Chrome)}: [\textbf{\color{purple}517$\uparrow$}, \textbf{\color{blue}1460$\downarrow$}, \textbf{\color{blue}1460$\downarrow$}, \textbf{\color{blue}502$\downarrow$}] \\ \hline
		macOS  & \textit{$TCP|322$ (APP)} or \textit{$TCP|49100$ (Chrome)}: [\textbf{\color{purple}517$\uparrow$}, \textbf{\color{blue}1412$\downarrow$}, \textbf{\color{blue}1412$\downarrow$}] \\ \hline
		android       & \textit{$TCP|322$ (APP)}: [\textbf{\color{purple}517$\uparrow$}, \textbf{\color{blue}3455$\downarrow$}]    \\ \hline
		iOS       & \textit{$TCP|49100$ (Safari)}: [\textbf{\color{purple}517$\uparrow$}, \textbf{\color{blue}3450$\downarrow$}]    \\ \hline
	\end{tabular}
	\label{tab:gamingManagementFlowSignatures}
\end{table}

According to prior research, flows with specific functionalities, such as console gaming \cite{SCMadanapalliPAM2022}, VoIP/video/file transfer/chat/browsing \cite{SRoyCC2022}, and encrypted web \cite{IAkbariMACS2021}, can be detected by the distinctive distribution or sequence of packet sizes they exhibit. 
Building on these findings, to detect platform and gameplay management flows without inspecting SNI, we leverage the sequence of payload sizes in the first few packets of a TCP flow, which contain predefined service requests after TCP three-way handshakes. 
As obtained from our training process, representative upstream and downstream packet payload sizes in gameplay management flows with their destination port numbers across different operating systems are provided in Table~\ref{tab:gamingManagementFlowSignatures}. The signatures for platform flows are not explicitly provided here for simplicity. 
With the precise specification of these signatures, we have achieved 100\% accuracy in detecting cloud gaming sessions during our lab evaluation.

It is important to note that our technique is designed for platform HTTPS flows and gameplay management TCP flows. The detection of gameplay UDP flows does not rely on service name signatures and is already encryption-agnostic, meaning it is not affected by future SNI encryption.

\subsection{Measuring Gameplay User Experience}\label{sec:QoEMetrics}
The user's experience in a cloud gameplay session is primarily determined by three factors: the synchronization speed of the user's mouse/keyboard input with the cloud platform (measured by client-platform latency in \S\ref{sec:clientPlatformLatency}), the smoothness of the streamed gaming scene (measured by video frame rate in \S\ref{sec:gamingVideoFrameRate}), and the clarity of the gaming graphics (indicated by graphic resolution in \S\ref{sec:gamingGraphicResolution}). 

In this section, we propose metrics to monitor these three key performance indicators derived from real-time volumetric statistics of gameplay session flows.
Our metrics are computed from transport-layer headers and packet sizes, therefore, are agnostic to the encryption of application-layer headers and payloads.

\subsubsection{Client-Platform Latency}\label{sec:clientPlatformLatency}
The first metric is client-platform latency, which represents the response time from the moment a gamer inputs commands with their keyboard or mouse till those commands are executed by the cloud platform.
As discussed in \S\ref{sec:GameSession}, there is a single gaming management flow over TCP that remains active throughout the entire gameplay session. This flow is detected by our methodology proposed in \S\ref{sec:detectingGameplaySession}.

The gaming management flow exhibits a constant packet rate of one pair of TCP packets every two seconds, where each pair consists of an upstream and downstream packet with matching \textbf{sequence} and \textbf{acknowledge numbers} in their TCP headers, as explained in \S\ref{sec:GamePlaySessionFlows}.

To measure the real-time latency experienced by the cloud gamer, we continuously monitor the arrival timestamps of each packet pair (identified by their sequence and acknowledge numbers) within the gaming management flow. Specifically, we track the timestamps $t_{up}$ and $t_{down}$ and calculate the latency $\Delta t$ between $t_{down}$ and $t_{up}$.

\renewcommand{\algorithmicrequire}{\textbf{Input:}}
\renewcommand{\algorithmicensure}{\textbf{Output:}}
\begin{algorithm}[t!]
	\caption{An algorithm for measuring the video frame rate (\ie frame count of an interval) of cloud gameplay using packet size patterns of downstream video flows.}
	\label{alg:frameRate}
	\begin{algorithmic}[1] 
		\REQUIRE \textit{packets} in gaming video flow; measurement interval $\Delta T$; payload size margin $\Delta size$
		\ENSURE measured \textit{frame\_count}
		
		\STATE $frame\_count \leftarrow 0$
		\STATE $t_{start} \leftarrow packets[0].arrival\_timestamp$
		\STATE $size_{max} \leftarrow packets[0].payload\_size$
		\STATE $flag_{max} \leftarrow FALSE$
		
		\FOR{$p$ \textbf{in} packets}
		\IF{$p.arrival\_timestamp > t_{start} + \Delta T$}
		\STATE \textbf{print} $frame\_count / \Delta T$
		\STATE $frame\_count \leftarrow 0$
		\ENDIF
		\IF{$p.payload\_size > size_{max}$}
		\STATE $size_{max} \leftarrow p.payload\_size$
		\STATE \textbf{continue}
		\ENDIF
		\IF{$p.payload\_size < size_{max} - \Delta size$}
		\IF{$flag_{max}$ is $TRUE$}
		\STATE $frame\_count \leftarrow frame\_count + 1$
		\STATE $flag_{max} \leftarrow FALSE$
		\ENDIF
		\STATE \textbf{continue}
		\ENDIF
		\STATE $flag_{max} \leftarrow TRUE$
		\ENDFOR
	\end{algorithmic}
\end{algorithm}

\subsubsection{Gaming Video Frame Rate}\label{sec:gamingVideoFrameRate}
The second metric we consider for user experience is the video frame rate being streamed to the cloud gamer. As discussed in \S\ref{sec:FrameRate}, a higher frame rate, such as 60fps, imposes stricter network requirements in terms of higher bandwidth and lower packet loss. In return, it offers the user a smoother gaming video experience.

To track this QoE metric, we leverage the periodic patterns of packet payload sizes observed in downstream video flows for both console applications and browsers during each gameplay session. These periodic patterns serve as a \textbf{direct} measure of the video frame rate, as explained in \S\ref{sec:FrameRate} and illustrated in Fig.~\ref{fig:videoFrame}. 
Considering that GeForce NOW offers frame rate options of 30fps and 60fps, we expect the observed count of periodical patterns per second to closely align with one of these two values.

The pseudocode block in Algorithm~\ref{alg:frameRate} shows our approach for measuring frame rate from downstream video flows.
The method takes as input the downstream packets in a gameplay video flow. It also requires an interval $\Delta T$ (set to 1 second in our implementation) that determines the frequency of measurements, and a payload size margin $\Delta size$ (fine-tuned using ground-truth sessions to 1 byte) that allows for variations in the payload size of full-size video packets in the flow.

To begin, the algorithm initializes four assisting variables from line \#1 to line \#4. Within the loop that processes the packet streams (line \#5), the measured frame rate per interval is reported and reset from line \#6 to line \#9. Additionally, the algorithm determines the full payload size of video packets in the monitored flow from line \#10 to line \#13.
From line \#14 to line \#21, the algorithm captures the stochastic periodical pattern observed in the packet payload sizes of a video flow. This pattern includes sequences of full-sized video packets followed by smaller-sized ones carrying frame markers and/or remaining data.

\subsubsection{Gaming Graphic Resolution}\label{sec:gamingGraphicResolution}
Our third QoE metric is graphic resolution, which represents the visual quality of the graphics being streamed to the cloud gamer by the cloud platform. The graphic resolution, along with the video frame rate, determine the bandwidth consumption of a downstream video flow, as discussed in \S\ref{sec:BandwidthConsumption}.
As just discussed, we can directly determine the current level of \textbf{video frame rate}. Therefore, to diagnose the real-time graphic resolution, we measure the current \textbf{bandwidth consumption} of the video flow and refer to the mapping provided in Table~\ref{tab:bitrate}. This mapping enables us to deduce the current graphic resolution based on the given throughput and frame rate.

\subsection{Discussion on Generalizability and Limitations}\label{sec:generalizabilityAndLimitation}
First, we have validated in our lab environment that the developed methods on cloud gaming detection, user setup identification, and gameplay QoE measurement can also be generalized to XBox Cloud Gaming platform with platform-specific signatures obtained from training traffic traces, including service domains, packet payload sizes, RTP port number, and flow volumetric criteria. Also, in the limited lab evaluation, the measurement methods for client-platform latency and video frame rate do not require signatures thus are directly applicable. However, the conclusions may need additional field evaluation to become fully valid.
Other popular platforms (listed in Table~\ref{tab:specification}) are not currently available in our region and we are not able to evaluate them in our lab. However, from a recent research \cite{JKyTNSM2023} studying the cloud gaming RTP flows, the platforms that are not covered in this study (\ie Playstation PLUS and Amazon Luna) share a common technological structure. Therefore, these can also be included in our method if lab setup for training trace collection is available. 

Second, due to potential future variations in the implementation of cloud gaming services, the alterations in the value of considered metrics (such as protocol type, volumetric statistics, and flow service domains) may diminish the effectiveness of a trained classification model. Consequently, retraining the model becomes necessary for optimal performance. Additionally, if significant modifications occur in the network anatomy of cloud gaming sessions (as briefly captured in Fig.~\ref{fig:gameplaySessionDetection}), such as the introduction of different categories of gameplay session flow types, adjustments to our model training process will be required.

Third, we acknowledge the limitations in our QoE metrics that are valuable topics for future research.	
The QoE metrics considered in this paper are directly related to network conditions, including graphic resolution and video frame rate impacted by available bandwidth and client-platform latency impacted by routing conditions. There are many other QoE metrics describing the capability of client devices and cloud servers \cite{HIqbalMACS2021}, such as processing delays for frame encoding, decoding, rendering and game engine processing.
While they are not directly impacted by the network conditions, it is worth investigating the correlation between those device-related QoE and network traffic characteristics such as signaling packets and inter-arrival timing, which may be predictable using statistical models. Also, the two gaming video metrics (\ie frame rate, resolution) may not always indicate true user experience as cloud gaming providers can dynamically adjust video settings for different in-game scenes \cite{network1030015,GBartolomeoCoNEXT2023}.

\section{Evaluation and Field Insights}\label{sec:evaluationAndFieldInsights}
We have implemented a fully-functional prototype of our cloud gaming detection and experience measurement framework in a large University campus with tens of thousands of students, including several hundred who reside in the dorms. Our system takes as input a raw feed of all traffic to/from the campus, obtained via optical taps on the fibres connecting the campus to the Internet. We begin by evaluating the accuracy of our system by playing cloud games from our lab on campus and comparing it to ground truth (\S\ref{sec:evaluation}), and then collect data in the wild over a 1-month period to demonstrate insights of interest to network operators (\S\ref{sec:deploymentInsights}).

\begin{figure*}[!t]
	\mbox{
		\subfigure[60FPS.]{
			{\includegraphics[width=0.3\textwidth]{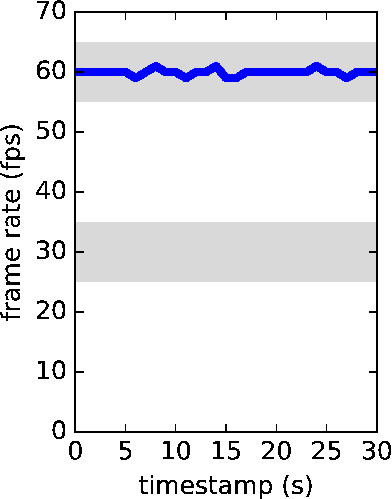}}\quad
			\label{fig:framerate60}
		}
		\hspace{-2mm}
		\subfigure[30FPS.]{
			{\includegraphics[width=0.3\textwidth]{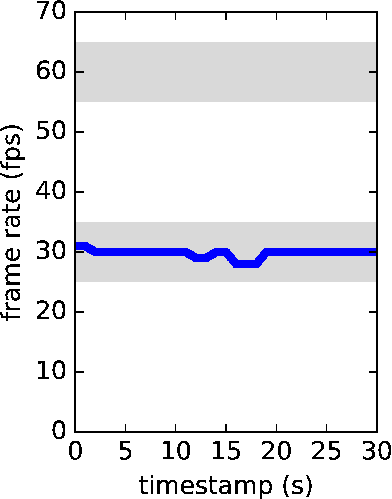}}\quad
			\label{fig:framerate30}
		}
		\hspace{-2mm}
		\subfigure[Non-steady.]{
			{\includegraphics[width=0.3\textwidth]{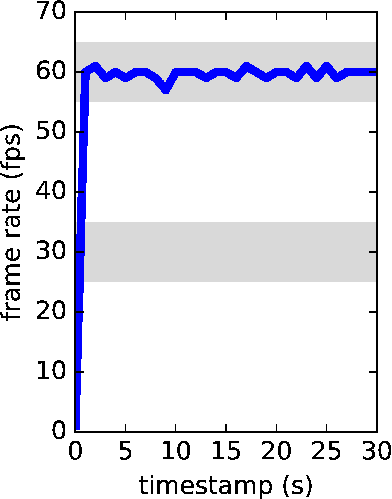}}\quad
			\label{fig:framerate30-start}
		}								
	}
	\caption{Example variations of the received video frame rate measured by our prototype.}
	\label{fig:evaluation}
\end{figure*}

\subsection{Lab Evaluation with Ground-Truth}\label{sec:evaluation}
Once our models were trained and deployed, we had volunteers play a total of 90 sessions of GeForce NOW gaming from our lab, and the ground truth they recorded was compared against the outputs reported from our system monitoring campus-wide traffic for the specific IP address of the lab devices. The sessions were designed to encompass various configurations, including user setups (PC app, mobile app, and browser), frame rates (60fps and 30fps), and video resolution bands (FHD, HD, and SD).

Our system correctly reported detection of all the cloud gaming sessions immediately upon commencement of actual gameplay, and the user setup (PC app, mobile app and browser play) was also identified with 100\% accuracy. The client-side latency recorded by our system (based on tracking TCP sequence numbers), the frame rates (measured from stochastic payload size patterns using the algorithms discussed in \S\ref{sec:FrameRate}), as well as the video resolution (inferred based on the real-time bandwidth usage and corresponding frame rate) corroborated accurately (\ie less than 2 fps deviation for frame rates and over 95\% accuracy for resolution) with the ground-truth collected on the client side, where traffic traces were recorded and analysed.

An interesting observation is that although the platform dynamically adjusts the frame rate to either 60fps or 30fps, in practice, the received frame rate may vary slightly (\eg within a range of 5fps) around the set rate. This variation is visually shown in Fig.~\ref{fig:framerate30} for the 30fps setting and Fig.~\ref{fig:framerate60} for the 60fps setting. Indeed there can be significant variations in the frame rate, particularly during the beginning of a gameplay session (as illustrated in Fig.~\ref{fig:framerate30-start}), or in situations where the network conditions are unstable.

\subsection{Field Deployment Insights}\label{sec:deploymentInsights}
We now present some insights obtained in the wild, as measured by our system over the entire month of May 2023 in the University campus network, that are of potential interest to ISPs on how user settings in GeForce NOW cloud gaming impact network bandwidth demand and end-user experience.
This equips ISPs to better understand their cloud gaming customer profiles, troubleshoot experience problems, and optimize network policies to better support cloud gaming flows using network slices, priority queues, and network APIs.

\begin{figure*}[t!]
	\begin{center}
		\mbox{
			\hspace{-4mm}
			\subfigure[Session.]{
				{\includegraphics[width=0.3\textwidth]{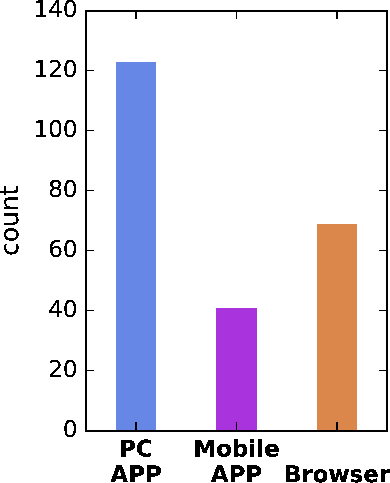}}\quad
				\label{fig:deploymentCount}
			}
			\subfigure[Play time.]{
				{\includegraphics[width=0.3\textwidth]{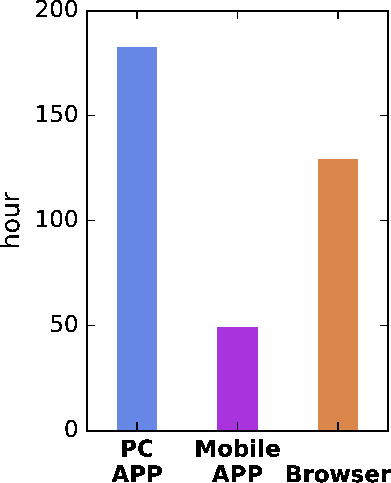}}\quad
				\label{fig:deploymentDuration}
			}
			\subfigure[Bandwidth.]{
				{\includegraphics[width=0.3\textwidth]{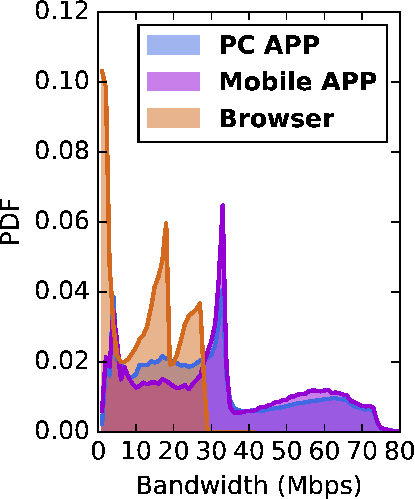}}\quad
				\label{fig:deploymentBandwidth}
			}			
		}
		\caption{GeForce NOW cloud gaming usage patterns from our 1-month campus deployment.} 
		\label{fig:deployment}
	\end{center}
\end{figure*}

\subsubsection{User Settings and Bandwidth Demand}
During the month of May, our system detected 233 cloud game sessions in the wild (\ie excluding the ones played by our volunteers), corresponding to 362 hours of playtime and 3.8 TB of data volume consumption. The majority of gameplay sessions (53\%) and gameplay time (51\%) were via the PC app, as shown in Figs.~\ref{fig:deploymentCount} and 
\ref{fig:deploymentDuration} respectively, followed by browsers (29\% of sessions and 36\% of total playtime), and finally the mobile app (17\% of sessions and 14\% of playtime). This is of relevance to ISPs because the bandwidth demanded by the gaming streams vary across these platforms -- Fig.~\ref{fig:deploymentBandwidth} shows the bandwidth distribution (measured over 1-second intervals) on the three platforms. The browser almost never exceeds 30 Mbps, while the PC and mobile app have significant durations with bandwidth demand in the range of 30-75 Mbps. While the underlying reason for this becomes clear when we examine video frame rates and resolutions next, visibility into app/browser mix may better equip ISPs in planning and provisioning their network capacity as cloud gaming grows.

\subsubsection{User Settings and Gaming Experience}\label{sec:gameExperience}
\begin{figure*}[t!]
	\begin{center}
		\mbox{
			\subfigure[Frame rate.]{
				{\includegraphics[width=0.28\textwidth]{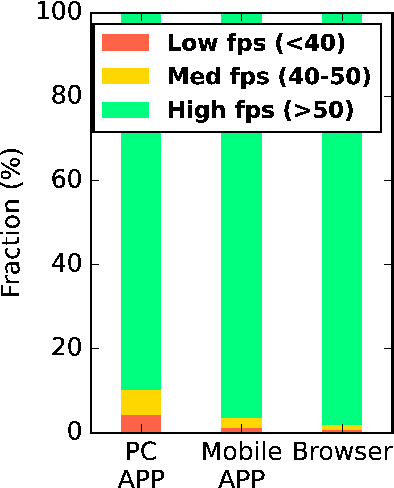}}\quad
				\label{fig:deploymentFrameRate}
			}
			\subfigure[Resolution.]{
				{\includegraphics[width=0.28\textwidth]{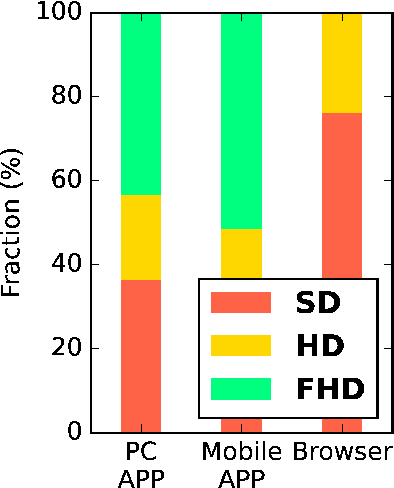}}\quad
				\label{fig:deploymentResolution}
			}				
			\subfigure[Latency (wired).]{
				{\includegraphics[width=0.28\textwidth]{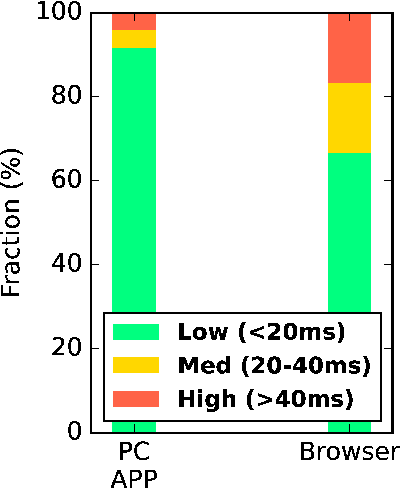}}\quad
				\label{fig:deploymentLatency}
			}				
		}
		\caption{GeForce NOW cloud gaming experience insights from our 1-month campus deployment.}
		\label{fig:deployment-experience}
	\end{center}
\end{figure*}

We saw in \S\ref{sec:evaluation} that frame rate can vary significantly, so for ease of depiction we group it into three bands: Low ($<$40 fps), Medium (40-50 fps) and High ($>$50 fps). Video resolution has already been banded into FHD, HD, and SD, as shown in Table~\ref{tab:bitrate}. For the three user settings (PC app, mobile app, and browser), we depict the percentage of time that the cloud game video operates at the three frame rate bands in Fig.~\ref{fig:deploymentFrameRate}, and the percentage of time the video renders in the three resolutions in Fig.~\ref{fig:deploymentResolution}. It is very interesting to note that browser gaming almost always has high fps but never goes to Full-HD resolution, whereas the PC app has FHD resolution 42\% of the time but drops fps to medium or low about 10\% of the time. The mobile app provides the best mix of fps and resolution overall, due to the advantage that it only has to work on a smaller screen. ISPs might use such visibility into the trade-offs on frame rates and resolution of cloud gaming sessions to better troubleshoot and support their customers on specific platforms.

Another interesting observation made by our system is in regards to the in-game latency across the platforms. To keep the comparison fair and not be influenced by wireless characteristics, we limited it to wired end-hosts (which are identifiable by IP address being on a different subnet to the WiFi network). Fig.~\ref{fig:deploymentLatency} shows that latency (averaged over 1-second intervals) stays under 20ms (the recommended threshold provided by Nvidia) 90\% of the time when played via the app, compared to 70\% when played via the browser. We believe this is because the app is better optimized for gaming than the browser -- indeed, we had shown in \S\ref{sec:FlowVolumetric} that the app maintains separate flows for user input and media, whereas the browser mixes all traffic into the one flow, which can lead to degraded jitter performance. ISPs may use such information to encourage customers who receive degraded experience to move from browser-based to app-based cloud gaming user setups.

\section{Related Work}\label{sec:relatedWork}
Cloud gaming has been the subject of prior studies that have explored various aspects \cite{WCaiAccess2016}, including cloud architecture \cite{RSheaNetwork2013,HChenTPDS2019,YHanTCC2022}, computing resource provisioning \cite{YLiHPDC2019,XZhangWC2019,GAMostafaJNCA2019}, gameplay video encoding \cite{GKIllahiTMCCA2020,ISlivarTMCCA2018}.
In chronological order, we discuss representative examples of prior works that have focused on enhancing gamer experience for different stakeholders, including cloud platform operators and mobile developers. It is important to note that our work specifically focuses on the cloud gaming experience for ISPs.
The authors of \cite{KChenToM2014} discussed several elements in the cloud platform that can have signifiant impact on user experiences if not well optimized.
Through a measurement study across Europe, the authors of \cite{TKamarainenMMSys2017} suggested optimal sever selection and control methods to achieve minimized latency for cloud gamers using mobile platforms.
In \cite{SSSabetMMVE2020}, the authors quantified user perception of graphic delays in different game genres (\eg card game, shooting, adventure, and racing) each with a unique sensitivity.
\textit{DECAF} \cite{HIqbalMACS2021} analysed the user-perceived experiences such as visual response received by user across different game genres. The authors highlight that networking issues like round trip delay, limited bandwidth, and packet losses could significantly degraded the user-perceived experience, which is quantitatively measured by our work using network QoE metrics. 
S. Bhuyan \textit{et al.} \cite{SBhuyanMACS2022} characterized the user-perceived performance (\eg frame rendering/decoding and hardware energy consumption) specifically for cloud gaming plays on mobile platforms over wireless networks.

As for network traffic analysis of cloud gaming platforms, M. Carrascosa \textit{et al.} \cite{MCarrascosaCC2022} identified generic traffic statistics (\eg utilized protocols, distribution of packet size, and packet inter-arrival time) of cloud gaming sessions on Google Stadia platform and their reactions toward sudden changes of network capacity.
X. Xu \textit{et al.} \cite{XXuIMC2022} measured the change in bandwidth of cloud gaming flows when competing with TCP Cubic and BBR flows.
Existing works such as \cite{JKyISCC2023,PGraffNOMS2023} developed NFV/SDN-based traffic processing systems to extract generic network attributes (\eg bitrate) of UDP flows to detect those carrying downstream video content of cloud games without considering other critical flows.
X. Marchal \textit{et al.} \cite{XMarchalJNSM2023,MCarrascosaCC2022} studied the network traffic patterns of cloud gaming platforms under constrained (cellular) network conditions.
The works described in \cite{JKyCNSM2022,JKyTNSM2023} analyzed performance anomalies (\eg channel degradation) of cloud gaming sessions served by 4G networks with time-series network KPIs.

In this paper, our objective is to address the lack of actionable visibility for ISPs. Compared to other works that analyze cloud gaming network traffic to investigate performance impact on RTP flows caused by suboptimal network conditions, we characterize network traffic patterns covering all service flows used in all stages of a cloud gaming session, carrying different gameplay functionalities, and across user setups. The comprehensive and unique insights obtained in our work are leveraged for gameplay detection, user setup identification, and measurement of cloud gaming user experiences.
\section{Conclusion}\label{sec:conclusion}

In this paper we developed a network traffic analysis method that provides ISPs visibility into cloud gaming sessions over their networks, specifically those served by Nvidia GeForce NOW platform. We first analyzed network traffic characteristics of cloud game sessions to establish benchmarks for critical service flows that exhibit unique patterns based on user setups and gameplay experiences. We then design a method to detect cloud game sessions across various user setups by stateful matching of service flows, classify critical gameplay flows using volumetric attributes, and track experience metrics (\ie client-platform latency, frame rate, and graphic resolution) on those flows. The method is prototyped in a large university network, evaluated using ground-truth sessions, and demonstrated for its operational insights from real-world cloud game sessions.\\

\noindent
{\textbf{Acknowledgements.}
We thank our shepherd Alessandro Finamore and the four anonymous reviewers for their insightful feedback. Funding for this project is provided by the Australian Government's Cooperative Research Centres Projects (CRC-P) Grant CRCPXIV000099.}

\newpage
\bibliographystyle{acm}
\bibliography{Reference}

\begin{thebibliography}{10}

\bibitem{IAkbariMACS2021}
{\sc Akbari, I., et~al.}
\newblock A look behind the curtain: Traffic classification in an increasingly
  encrypted web.
\newblock {\em Proc. ACM Meas. Anal. Comput. Syst.\/} (Feb 2021).

\bibitem{Luna}
{\sc {Amazon}}.
\newblock {Luna}.
\newblock \url{https://www.amazon.com/luna/landing-page}, 2023.
\newblock Accessed: 2023-01-12.

\bibitem{GBartolomeoCoNEXT2023}
{\sc Bartolomeo, G., Cao, J., Su, X., and Mohan, N.}
\newblock Characterizing distributed mobile augmented reality applications at
  the edge.
\newblock In {\em Proc. ACM CoNEXT\/} (Paris, France, Dec 2023).

\bibitem{SBhuyanMACS2022}
{\sc Bhuyan, S., Zhao, S., Ying, Z., Kandemir, M.~T., and Das, C.~R.}
\newblock End-to-end characterization of game streaming applications on mobile
  platforms.
\newblock {\em Proc. ACM Meas. Anal. Comput. Syst.\/} (Feb 2022).

\bibitem{Economist2022}
{\sc {Business}}.
\newblock {Microsoft, Activision Blizzard and the Future of Gaming}.
\newblock {\em The Economist\/} (Nov 2022).

\bibitem{WCaiAccess2016}
{\sc Cai, W., Shea, R., Huang, C.-Y., Chen, K.-T., Liu, J., Leung, V. C.~M.,
  and Hsu, C.-H.}
\newblock A survey on cloud gaming: Future of computer games.
\newblock {\em IEEE Access 4\/} (2016), 7605--7620.

\bibitem{MCarrascosaCC2022}
{\sc Carrascosa, M., and Bellalta, B.}
\newblock Cloud-gaming: Analysis of google stadia traffic.
\newblock {\em Computer Communications 188\/} (2022), 99--116.

\bibitem{HChenTPDS2019}
{\sc Chen, H., Zhang, X., Xu, Y., Ren, J., Fan, J., Ma, Z., and Zhang, W.}
\newblock T-gaming: A cost-efficient cloud gaming system at scale.
\newblock {\em IEEE Transactions on Parallel and Distributed Systems 30}, 12
  (2019), 2849--2865.

\bibitem{KChenToM2014}
{\sc Chen, K.-T., Chang, Y.-C., Hsu, H.-J., Chen, D.-Y., Huang, C.-Y., and Hsu,
  C.-H.}
\newblock On the quality of service of cloud gaming systems.
\newblock {\em IEEE Transactions on Multimedia 16}, 2 (2014), 480--495.

\bibitem{SNI}
{\sc {CloudFlare}}.
\newblock {What is SNI? How TLS Server Name Indication Works}.
\newblock \url{https://www.cloudflare.com/en-gb/learning/ssl/what-is-sni/},
  2022.
\newblock Accessed: 2023-01-12.

\bibitem{network1030015}
{\sc Di~Domenico, A., Perna, G., Trevisan, M., Vassio, L., and Giordano, D.}
\newblock A network analysis on cloud gaming: Stadia, geforce now and psnow.
\newblock {\em Network 1}, 3 (2021), 247--260.

\bibitem{GAMostafaJNCA2019}
{\sc Ghobaei-Arani, M., Khorsand, R., and Ramezanpour, M.}
\newblock An autonomous resource provisioning framework for massively
  multiplayer online games in cloud environment.
\newblock {\em Journal of Network and Computer Applications 142\/} (2019),
  76--97.

\bibitem{GitnuxBlog}
{\sc {Gitnux}}.
\newblock {Cloud Gaming Services: A Look at the Latest Statistics}.
\newblock
  \url{https://blog.gitnux.com/cloud-gaming-services-statistics/\#content},
  2023.
\newblock Accessed: 2023-06-26.

\bibitem{PGraffNOMS2023}
{\sc Graff, P., Marchal, X., Cholez, T., Mathieu, B., and Festor, O.}
\newblock Efficient identification of cloud gaming traffic at the edge.
\newblock In {\em Proc. IEEE/IFIP Network Operations and Management
  Symposium\/} (May 2023).

\bibitem{YHanTCC2022}
{\sc Han, Y., Guo, D., Cai, W., Wang, X., and Leung, V. C.~M.}
\newblock Virtual machine placement optimization in mobile cloud gaming through
  qoe-oriented resource competition.
\newblock {\em IEEE Transactions on Cloud Computing 10}, 3 (2022), 2204--2218.

\bibitem{GKIllahiTMCCA2020}
{\sc Illahi, G.~K., Gemert, T.~V., Siekkinen, M., Masala, E., Oulasvirta, A.,
  and Yl\"{a}-J\"{a}\"{a}ski, A.}
\newblock Cloud gaming with foveated video encoding.
\newblock {\em ACM Trans. Multimedia Comput. Commun. Appl. 16}, 1 (Feb 2020).

\bibitem{HIqbalMACS2021}
{\sc Iqbal, H., Khalid, A., and Shahzad, M.}
\newblock Dissecting cloud gaming performance with decaf.
\newblock {\em Proc. ACM Meas. Anal. Comput. Syst.\/} (Dec 2021).

\bibitem{TKamarainenMMSys2017}
{\sc K\"{a}m\"{a}r\"{a}inen, T., Siekkinen, M., Yl\"{a}-J\"{a}\"{a}ski, A.,
  Zhang, W., and Hui, P.}
\newblock A measurement study on achieving imperceptible latency in mobile
  cloud gaming.
\newblock In {\em Proc. ACM MMSys\/} (Taipei, Taiwan, Jun 2017).

\bibitem{CMS}
{\sc {Kinsta}}.
\newblock {What is a Content Management System (CMS)?}
\newblock \url{https://kinsta.com/knowledgebase/content-management-system/},
  2022.
\newblock Accessed: 2023-01-12.

\bibitem{JKyISCC2023}
{\sc Ky, J., Graff, P., Mathieu, B., and Cholez, T.}
\newblock A hybrid p4/nfv architecture for cloud gaming traffic detection with
  unsupervised ml.
\newblock In {\em Proc. IEEE Symposium on Computers and Communications\/} (Los
  Alamitos, CA, USA, jul 2023).

\bibitem{JKyCNSM2022}
{\sc Ky, J.~R., Mathieu, B., Lahmadi, A., and Boutaba, R.}
\newblock Assessing unsupervised machine learning solutions for anomaly
  detection in cloud gaming sessions.
\newblock In {\em Proc. IEEE International Conference on Network and Service
  Management\/} (Thessaloniki, Greece, Oct 2022).

\bibitem{JKyTNSM2023}
{\sc Ky, J.~R., Mathieu, B., Lahmadi, A., and Boutaba, R.}
\newblock Ml models for detecting qoe degradation in low-latency applications:
  A cloud-gaming case study.
\newblock {\em IEEE Transactions on Network and Service Management\/} (2023).

\bibitem{YLiHPDC2019}
{\sc Li, Y., Shan, C., Chen, R., Tang, X., Cai, W., Tang, S., Liu, X., Wang,
  G., Gong, X., and Zhang, Y.}
\newblock Gaugur: Quantifying performance interference of colocated games for
  improving resource utilization in cloud gaming.
\newblock In {\em Proc. ACM HPDC\/} (Phoenix, AZ, USA, Jun 2019).

\bibitem{SLiuMWUT2023}
{\sc Liu, S., Mangla, T., Shaowang, T., Zhao, J., Paparrizos, J., Krishnan, S.,
  and Feamster, N.}
\newblock Amir: Active multimodal interaction recognition from video and
  network traffic in connected environments.
\newblock {\em Proc. ACM Interact. Mob. Wearable Ubiquitous Technol.\/} (mar
  2023).

\bibitem{comcast}
{\sc Livingood, J.}
\newblock {Comcast Kicks Off Industry's First Low Latency DOCSIS Field Trials}.
\newblock
  \url{https://corporate.comcast.com/stories/comcast-kicks-off-industrys-first-low-latency-docsis-field-trials},
  2023.
\newblock Accessed: 2023-06-26.

\bibitem{SCMadanapalliPAM2022}
{\sc Madanapalli, S.~C., Gharakheili, H.~H., and Sivaraman, V.}
\newblock {Know Thy Lag: In-Network Game Detection And Latency Measurement}.
\newblock In {\em Proc. PAM\/} (Apr 2022).

\bibitem{SCMadanapalliIWQoS2021}
{\sc Madanapalli, S.~C., Mathai, A., Gharakheili, H.~H., and Sivaraman, V.}
\newblock Reclive: Real-time classification and qoe inference of live video
  streaming services.
\newblock In {\em Proc. IEEE/ACM IWQOS\/} (2021).

\bibitem{XMarchalJNSM2023}
{\sc Marchal, Xavierl, G. P. K. J.~R., Cholez, T., Tuffin, S., Mathieu, B., and
  Festor, O.}
\newblock An analysis of cloud gaming platforms behaviour under synthetic
  network constraints and real cellular networks conditions.
\newblock {\em Journal of Network and Systems Management\/} (2023).

\bibitem{ClougGamingMarket}
{\sc {Markets and Markets}}.
\newblock {Cloud Gaming Market by Offiering, Device Type, Solution, Game Type,
  Region -- Global Forecast to 2024}.
\newblock \url{https://bit.ly/3AyEjio}, 2019.
\newblock Accessed: 2024-04-27.

\bibitem{XBoxCloudGame}
{\sc {Microsoft}}.
\newblock {XBox Cloud Gaming (Beta)}.
\newblock \url{https://www.xbox.com/en-us/play}, 2023.
\newblock Accessed: 2023-01-12.

\bibitem{Economist2023}
{\sc {News}}.
\newblock {The Future of Video Games}.
\newblock {\em The Economist\/} (Mar 2023).

\bibitem{GeForceNOW}
{\sc {Nvidia}}.
\newblock {GeForce NOW}.
\newblock \url{https://www.nvidia.com/en-au/geforce-now/}, 2023.
\newblock Accessed: 2023-01-12.

\bibitem{NividiaPortMapping}
{\sc {Nvidia Support}}.
\newblock How can i reduce lag or improve streaming quality when using geforce
  now?
\newblock \url{bit.ly/45TOmfR}, 2022.
\newblock Accessed: 2022-12-12.

\bibitem{NividiaPortWebRTC}
{\sc {Nvidia Support}}.
\newblock {WebRTC Browser Client}.
\newblock \url{bit.ly/3LhOPAj}, 2022.
\newblock Accessed: 2022-12-14.

\bibitem{JRietveld2023}
{\sc Rietveld, J.}
\newblock Microsoft and activision: the big questions that will decide whether
  the us\$68 billion deal goes ahead.
\newblock {\em The Conversation\/} (Jul 2023).

\bibitem{SRoyCC2022}
{\sc Roy, S., Shapira, T., and Shavitt, Y.}
\newblock Fast and lean encrypted internet traffic classification.
\newblock {\em Computer Communications 186\/} (2022), 166--173.

\bibitem{SSSabetMMVE2020}
{\sc Sabet, S.~S., Schmidt, S., Zadtootaghaj, S., Griwodz, C., and M\"{o}ller,
  S.}
\newblock Delay sensitivity classification of cloud gaming content.
\newblock In {\em Proceedings of the 12th ACM International Workshop on
  Immersive Mixed and Virtual Environment Systems\/} (Istanbul, Turkey, 2020).

\bibitem{RTPRFC}
{\sc Schulzrinne, H., Casner, S., Frederick, R., and Jacobson, V.}
\newblock {{RTP: A Transport Protocol for Real-Time Applications}}.
\newblock {RFC} 3550, Jul 2003.

\bibitem{sharma2023estimating}
{\sc Sharma, T., Mangla, T., Gupta, A., Jiang, J., and Feamster, N.}
\newblock Estimating webrtc video qoe metrics without using application
  headers, 2023.

\bibitem{RSheaNetwork2013}
{\sc Shea, R., Liu, J., Ngai, E. C.-H., and Cui, Y.}
\newblock Cloud gaming: architecture and performance.
\newblock {\em IEEE Network 27}, 4 (2013), 16--21.

\bibitem{ISlivarTMCCA2018}
{\sc Slivar, I., Suznjevic, M., and Skorin-Kapov, L.}
\newblock Game categorization for deriving qoe-driven video encoding
  configuration strategies for cloud gaming.
\newblock {\em ACM Trans. Multimedia Comput. Commun. Appl.\/} (jun 2018).

\bibitem{PlayStationNOW}
{\sc {Sony Interactive Entertainment}}.
\newblock {PlayStation Now}.
\newblock \url{https://www.playstation.com/en-us/ps-now/}, 2023.
\newblock Accessed: 2023-04-18.

\bibitem{BSpangBS2020}
{\sc Spang, B., Walsh, B., Huang, T.-Y., Rusnock, T., Lawrence, J., and
  McKeown, N.}
\newblock Buffer sizing and video qoe measurements at netflix.
\newblock In {\em Proceedings of the 2019 Workshop on Buffer Sizing\/} (2020).

\bibitem{NWehnerPER2021}
{\sc Wehner, N., Seufert, M., Schuler, J., Wassermann, S., Casas, P., and
  Hossfeld, T.}
\newblock Improving web qoe monitoring for encrypted network traffic through
  time series modeling.
\newblock {\em SIGMETRICS Perform. Eval. Rev.\/} (may 2021).

\bibitem{XXuIMC2022}
{\sc Xu, X., and Claypool, M.}
\newblock Measurement of cloud-based game streaming system response to
  competing tcp cubic or tcp bbr flows.
\newblock In {\em Proc. ACM Internet Measurement Conference\/} (Nice, France,
  2022).

\bibitem{XZhangWC2019}
{\sc Zhang, X., Chen, H., Zhao, Y., Ma, Z., Xu, Y., Huang, H., Yin, H., and Wu,
  D.~O.}
\newblock Improving cloud gaming experience through mobile edge computing.
\newblock {\em IEEE Wireless Communications 26}, 4 (2019), 178--183.

\end{thebibliography}

\appendix

\vspace{-8mm}
\section{Flow Profiles of GeForce NOW Cloud Game Sessions across User Setups}\label{sec:AppendixGFNFLowProfile}
\vspace{-4mm}
 \begin{figure*}[t!]
	\begin{center}
		\mbox{
			\subfigure[Browser.]{
				{\includegraphics[width=0.45\textwidth]{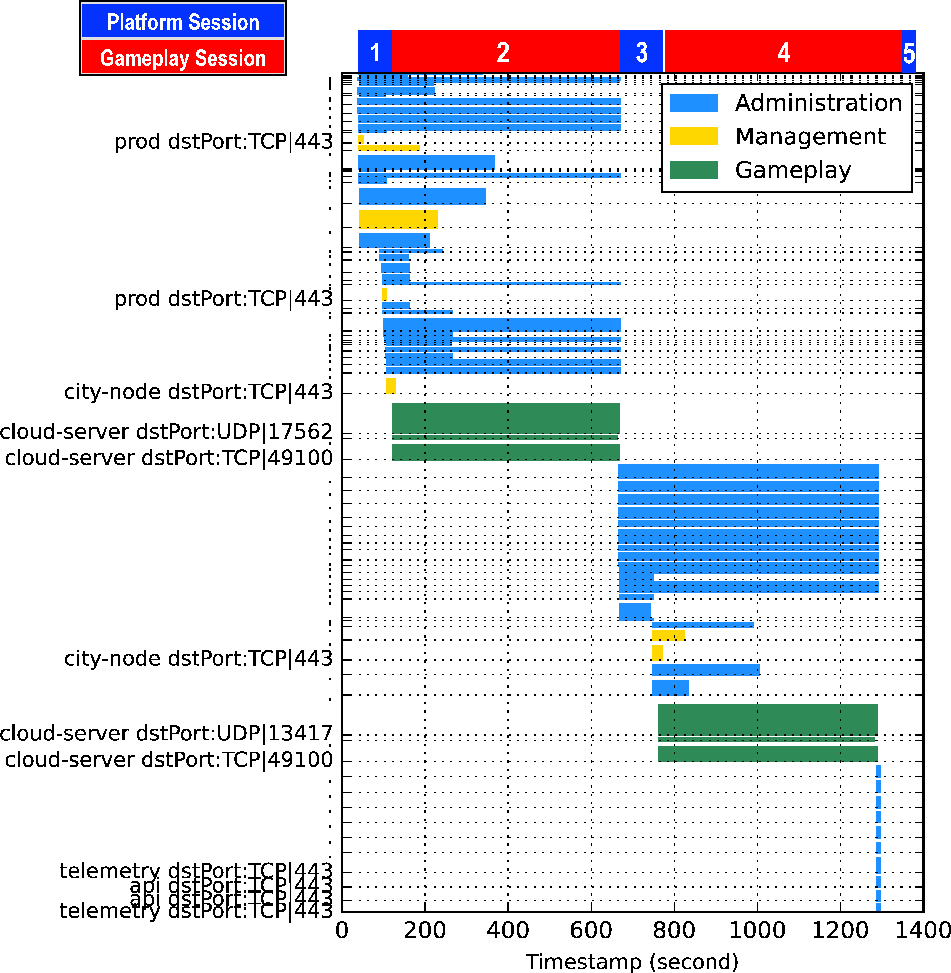}}\quad
				\label{fig:FlowProfileChrome}
			}
			\subfigure[Mobile console application.]{
				{\includegraphics[width=0.45\textwidth]{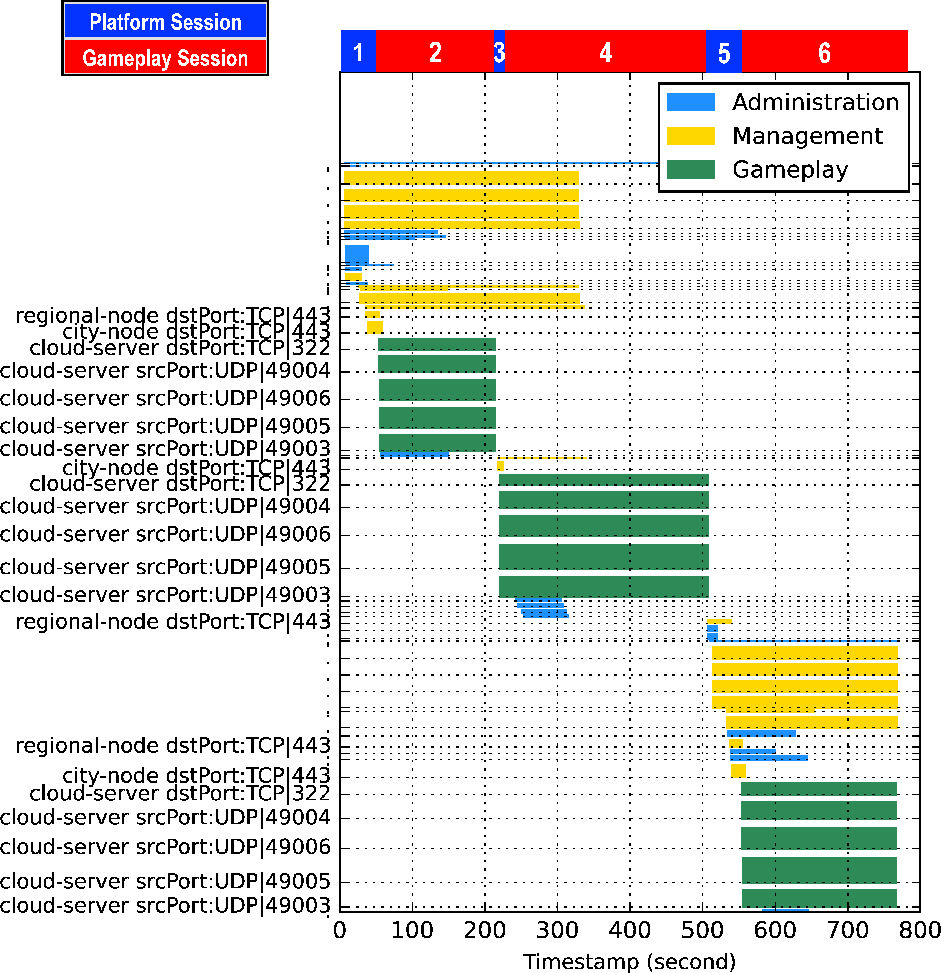}}\quad
				\label{fig:FlowProfileAndroid}
			}
		}
			\vspace{-5mm}
		\caption{Flow profiles of example gameplay via different user setups. Service prefixes and port numbers of representative flows are shown by their respective y-ticks, and the throughput of each flow is shown by its thickness (normalized by logarithmic functions).}
		\label{fig:flowProfileAppendix}
	\end{center}
	\vspace{-6mm}
\end{figure*}

 \begin{figure*}[t!]
	\begin{center}
		\mbox{
			\subfigure[XBox hardware console.]{
				{\includegraphics[width=0.3\textwidth]{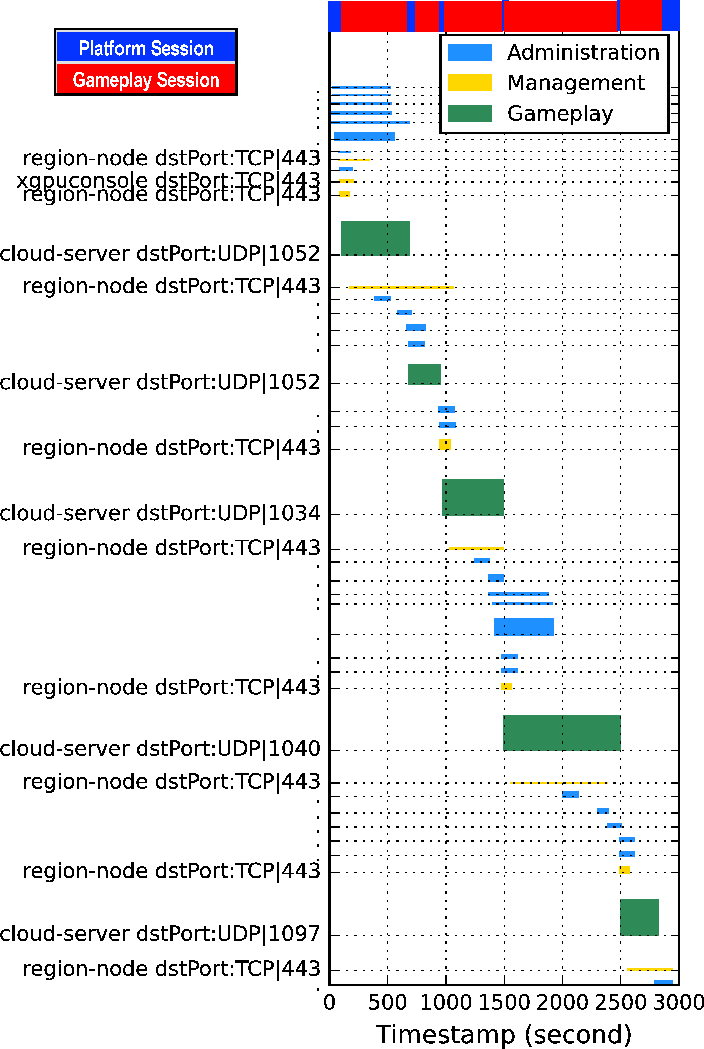}}\quad
				\label{fig:FlowProfileXBoxConsole}
			}
			\subfigure[PC browser.]{
				{\includegraphics[width=0.3\textwidth]{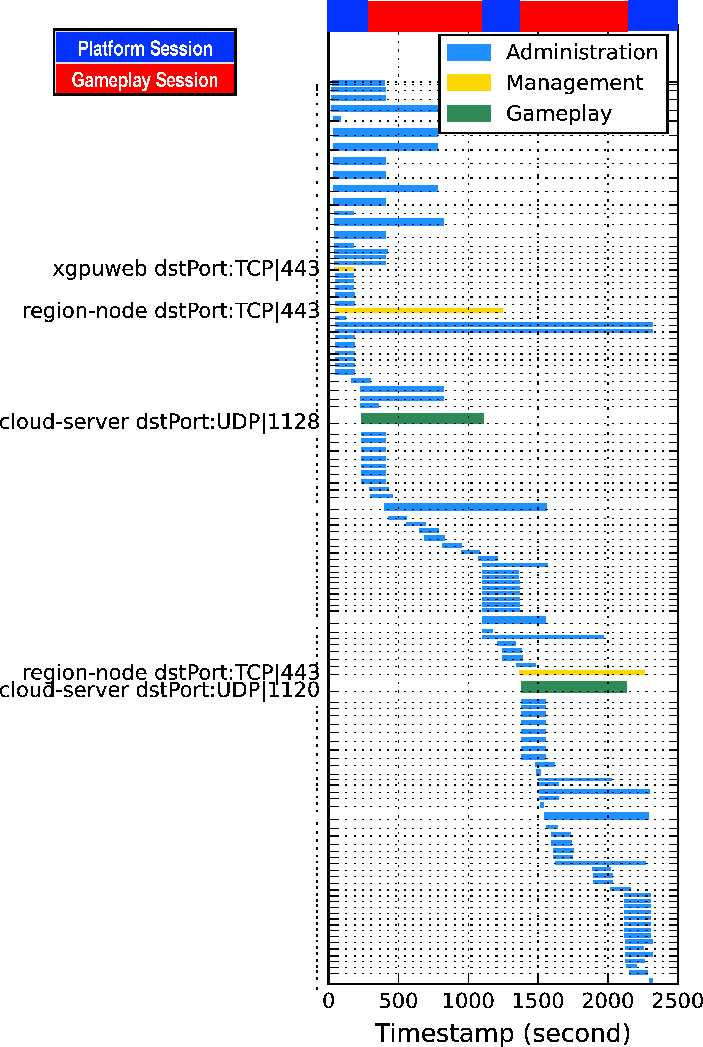}}\quad
				\label{fig:FlowProfileXBoxPC}
			}
			\subfigure[Mobile browser.]{
				{\includegraphics[width=0.3\textwidth]{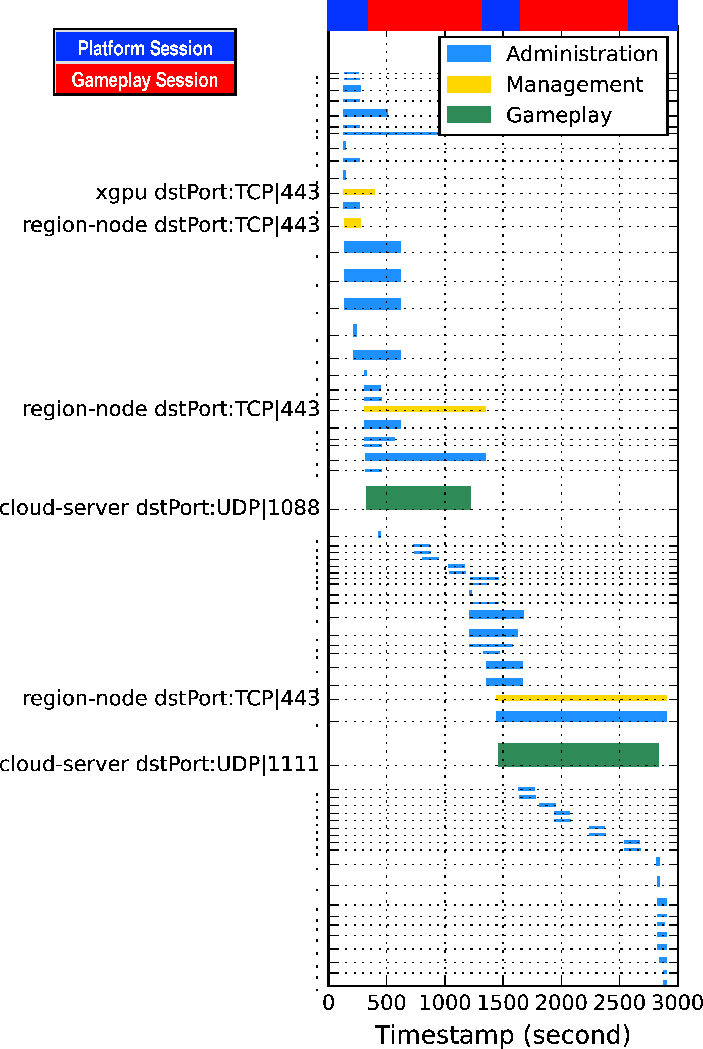}}\quad
				\label{fig:FlowProfileXBoxMobile}
			}
		}
		\vspace{-5mm}
		\caption{Flow profiles of XBox Cloud Gaming sessions via different user setups. Service prefixes and port numbers of representative flows are shown by their respective y-ticks, and the throughput of each flow is shown by its thickness (normalized by logarithmic functions).}
		\label{fig:flowProfileXBoxAppendix}
	\end{center}
	\vspace{-10mm}
\end{figure*}

As complimentary to the flow profiles for cloud game sessions via desktop console application given in Fig.~\ref{fig:FlowProfileGFN}, the ones via browser and mobile console applications are shown in  Fig.~\ref{fig:flowProfileAppendix}. Comparing the three figures, they all have different usage of service flows for platform administration and game session management purposes. Sessions via both mobile and desktop console applications have five gameplay flows each with a unique functionality, whereas those via browsers only have two gameplay flows, one for management, and one for combined user input and game media streaming.

\vspace{-4mm}
\section{Network Traffic Characteristics of XBox Cloud Gaming Sessions}\label{sec:AppendixXBoxFlowProfile}
\vspace{-4mm}
We have conducted similar lab experiments on another cloud gaming platform Microsoft's XBox Cloud Gaming that is currently available in Australia.
We discuss flow profiles of gameplay sessions on XBox Cloud Gaming platform and the applicability of our developed methods.
 
 Fig.~\ref{fig:flowProfileXBoxAppendix} shows the usage of service flows during cloud gaming sessions on XBox Cloud Gaming platform accessed via three supported user setups including XBox hardware console (Fig.~\ref{fig:FlowProfileXBoxConsole}), PC browser (Fig.~\ref{fig:FlowProfileXBoxPC}), and mobile browser (Fig.~\ref{fig:FlowProfileXBoxMobile}), respectively.
 Compared to the evolution of service flows in Nvidia GeForce NOW we discussed in \S\ref{sec:TrafficAnalysis}, we observe similar insights.
 First of all, the purposes of flows are also categorized into platform administration, platform management, and gameplay; second, prior to each gameplay session, platform management flows (annotated as ``\textit{regional-node}'') are started to check current network connectivity and select appropriate cloud server; third, RTP flows that are destinated to a certain range of port numbers (\eg $UDP|1040$ to $UDP|1190$) on the cloud server are used for gaming media and user input.
 Similar to GeForce NOW, the services being accessed in platform sessions of XBox Cloud Gaming vary across user setup. As shown by example service names  in Fig.~\ref{fig:FlowProfileXBoxConsole}, \ref{fig:FlowProfileXBoxPC} and \ref{fig:FlowProfileXBoxMobile}, sessions accessed by XBox hardware console, PC browser and mobile browser uses different graphic services namely \textit{xgpuconsole}, \textit{xgpuweb} and \textit{xgpu}, respectively.
 
 Apart from specific domain name (\eg XBox Cloud Gaming uses \textit{xboxlive.com} as its gameplay domain while GeForce NOW uses \textit{nvidiagrid.net}) and range of service port numbers used on the cloud server that are different between XBox and GFN, we observed that XBox uses a single RTP flow for both streaming media and user input even on its native hardware console, while GFN uses separate RTP flows each only carry one type of traffic for sessions from console application.
 
 It is not surprising to observe the above commonalities as cloud gaming platforms are built on similar technological paradigms and communication protocols. Therefore, our methods in detecting cloud gaming session, identifying user setup, and measuring gameplay QoE metrics are evidently applicable to XBox Cloud Gaming and other platforms with similar underlying structures.
  
\vspace{-4mm}
\section{Charts Illustrating Classification of Gameplay Flows using Heuristically Simplified Criteria}\label{sec:AppendixGameplayFlowClassification}
\begin{figure*}[h]
	\vspace{-5mm}
	{\includegraphics[width=\textwidth]{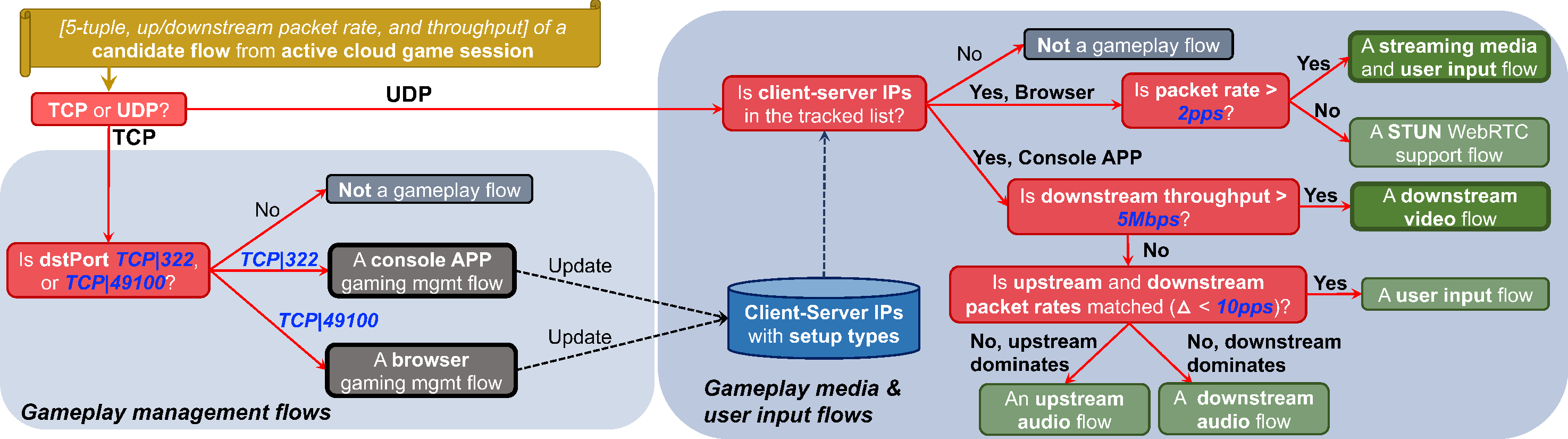}}
		\vspace{-7mm}
	\caption{Illustrative process for the classification of Nvidia's \textbf{GeForce NOW} gameplay session flows, wherein the criteria obtained from our training process is annotated with \textbf{\color{blue}blue text}.}
	\label{fig:gameplaySessionDetection-GFN}
	\vspace{-7mm}
\end{figure*}

\begin{figure*}[h]
	\vspace{-5mm}
	{\includegraphics[width=\textwidth]{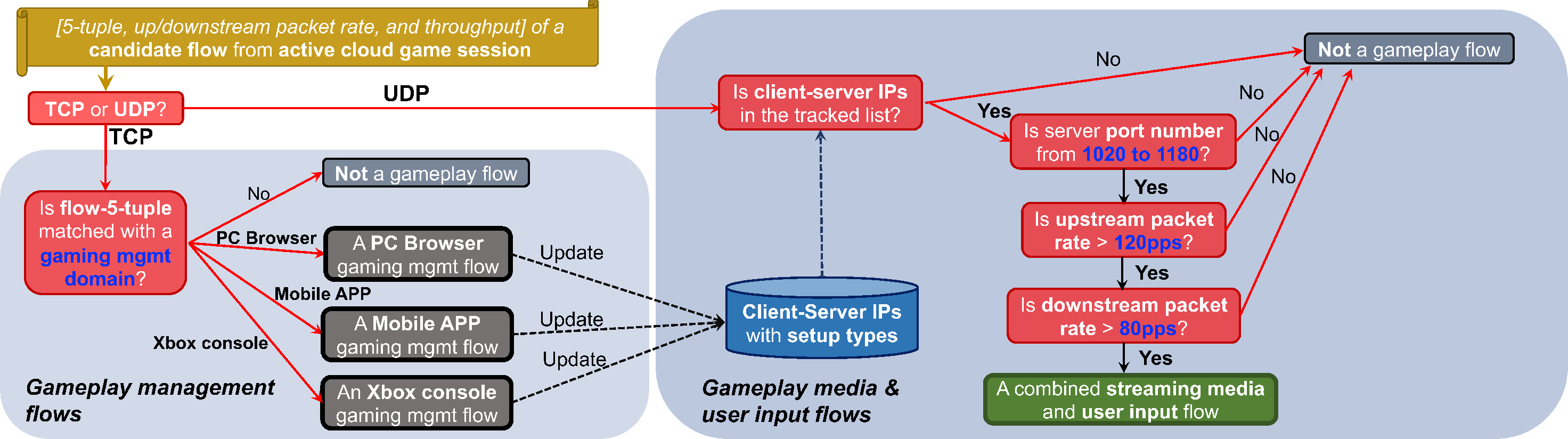}}
		\vspace{-7mm}
	\caption{Illustrative process for the classification of Microsoft's \textbf{XBox Cloud Gaming} gameplay session flows, wherein the criteria obtained from our training process is annotated with \textbf{\color{blue}blue text}.}
	\label{fig:gameplaySessionDetection-XBox}
	\vspace{-7mm}
\end{figure*}

\vspace{-1mm}
Fig.~\ref{fig:gameplaySessionDetection-GFN} and \ref{fig:gameplaySessionDetection-XBox} visually show the gameplay session flow classification processes for GeForce NOW and Xbox Cloud Gaming derived from our generic process shown in Fig.~\ref{fig:gameplaySessionDetection} with heuristically simplified criteria obtained from our training process on ground-truth traffic traces.

\end{document}